
\magnification\magstep1
\tolerance=1600
\parskip=0pt
\baselineskip= 6 true mm

\def\ft#1#2{{\textstyle{{#1}\over{#2}}}}
\def\a{\alpha}
\def\b{\beta}
\def\g{\gamma}\def\G{\Gamma}
\def\d{\delta}
\def\e{\epsilon}

\def\k{{\kappa}}

\def\l{\lambda}
\def\m{\mu}
\def\f{\phi}
\def\n{\nu}
\def\p{\psi}
\def\r{\rho}
\def\s{\sigma}

\def\x{\chi}
\def\o{\omega}\def\O{\Omega}
\def\var{\varepsilon}

\def\slash{\llap /}
\def\lagr{{\cal L}}
\def\pa{\partial}

\def\Gnul{{\buildrel \circ\over \Gamma}{}}
\def\cstr#1#2#3{{f^{\,#2}_{#1#3}}}

\def\cV{{\cal V}}
\def\half{{\textstyle {1 \over 2}}}
\def\quart{{\textstyle {1 \over 4}}}
\def\GIJ{\Gamma^{IJ}}
\def\HA{\Gamma^{(2)}}
\def\HB{\Gamma^{(4)}}
\def\HC{\Gamma^{(6)}}

{\nopagenumbers

\vglue .5truecm
\rightline{CERN-TH.6612/92}
\rightline{THU-92-18}

\vskip 1.1truecm
\centerline{{\bf LOCALLY SUPERSYMMETRIC $D=3$ NON-LINEAR SIGMA
MODELS}}

\vskip 1.3truecm

\centerline{B.~de Wit\footnote{*}{Permanent address:
Institute for Theoretical Physics, University of
Utrecht,\hfil\break\indent The Netherlands.},
A.K.~Tollst\'en }
\vskip .6truemm
\centerline{Theory Division, CERN}
\centerline{CH-1211, Geneva 23, Switzerland}
\vskip .2truecm
\centerline{and}
\vskip .2truecm
\centerline{H.~Nicolai}
\vskip .6truemm
\centerline{II. Institute for Theoretical Physics, University of
Hamburg}
\centerline{Luruper Chaussee 149, 2000 Hamburg 50, F.R.G. }
\vskip .5true cm

\midinsert
\baselineskip=5true mm
\centerline{\bf Abstract}\vskip .1 true cm
\narrower
\noindent We study non-linear sigma models with $N$
local supersymmetries in three space-time dimensions. For
$N=1$ and 2 the target space of these models is Riemannian or
K\"ahler, respectively. All $N>2$ theories
are associated with Einstein spaces. For $N=3$ the target space is
quaternionic, while for $N=4$ it generally decomposes
into two separate quaternionic spaces,
associated with inequivalent supermultiplets. For $N=5,6,8$ there
is a unique (symmetric) space for any given number of supermultiplets.
Beyond that there are only theories based on a
single supermultiplet for $N=9,10,12$ and 16, associated with
coset spaces with the exceptional isometry groups $F_{4(-20)}$,
$E_{6(-14)}$,  $E_{7(-5)}$ and $E_{8(+8)}$, respectively. For
$N=3$  and $N\geq 5$ the $D=2$ theories obtained by dimensional
reduction are two-loop finite.
\endinsert

\footnote{}{}
\footnote{}{CERN-TH.6612/92}
\footnote{}{THU-92-18}
\footnote{}{August 1992}

\vfill\eject
}
\pageno=1

\beginsection 1. Introduction

For space-time dimensions $D\geq 4$ a large variety of locally
supersymmetric theories has been explored, both with and without
conformal invariance [1]. For $D=2$ conformally
invariant theories have been studied extensively. In contrast,
only very few models have been worked out for $D=3$.
Nevertheless, gravity and supergravity in three dimensions
are of interest in their own right. As is well known,
three-dimensional field theories have a number
of unique features. For instance, massless states do not carry
helicity, so that the associated degrees of freedom
can generally be described by scalar fields. Pure gravity and
supergravity are topological theories and do not give rise to
physical (i.e. propagating) degrees of freedom. Apart from conical
singularities at the location of matter sources, space-time is flat.
Notwithstanding this fact, classical gravity in three dimensions
exhibits many intriguing properties [2]. More recently,
pure quantum gravity in three dimensions has been reformulated
as a Chern-Simons gauge theory and shown to be solvable
in the sense that the quantum constraints
(i.e. the Wheeler-DeWitt equation, in particular) can be solved
exactly [3]. In addition, genuine observables (\`a la Dirac)
can be constructed, in contrast to four-dimensional canonical
gravity, where the construction of observables remains an unsolved
problem even at the classical level. Moreover, the three-dimensional
theory is especially amenable to a reformulation in terms of the new
canonical variables proposed in [4] (see also [5]
for a clear discussion); the exact solvability of pure quantum
gravity in this approach has been demonstrated in [6].
A recent treatment of pure and matter-coupled supergravity in this
framework can be found in [7]. Although many open questions
remain, it should be clear from these remarks that three-dimensional
gravity and supergravity can teach us a lot about quantum gravity
in general, and that the models considered here, at the very least,
can serve as non-trivial toy models.

A further motivation for studying three-dimensional supergravity is
the important role it plays in the construction of
two-dimensional supergravity theories via dimensional reduction.
These dimensionally reduced theories have a number of remarkable
properties; in particular, they possess infinite-dimensional
symmetries acting on the space of solutions of the non-linear field
equations [8,9,10]. For supergravity, these symmetries
merge with the so-called ``hidden symmetries" of supergravity.
All these models are classically integrable in the sense
that they admit linear systems for their non-linear field
equations [9,10]. The belief that this classical symmetry
structure should play an important role for the quantum theory
was one of the main motivations for a recent
investigation of the quantum divergences of these
two-dimensional supergravity theories [11], which showed that
for sufficiently high $N$ (the number of independent
supersymmetries) these models were two-loop finite. In order to
appreciate the relevance of this result, it is important to
understand the uniqueness of these theories. In [11] the
calculations were based on the conjectured structure of
non-linear sigma models coupled to $D=3$ supergravity with
homogeneous target spaces, as they were known or expected to
arise by dimensional reduction from extended supergravity in four
space-time dimensions, but to date only a few of these models
have been worked out explicitly [12,7].

The present paper aims at filling this gap and gives a
complete classification of non-linear sigma models coupled to
extended supergravity in three space-time dimensions.
For rigidly supersymmetric non-linear sigma models, this analysis
is almost identical to the $D=2$ case [13]. There it was
established that $N$-extended supersymmetric sigma models
require the presence of $N-1$ complex structures in the target
space. It turns out that non-linear sigma models based on
irreducible target spaces can have at most $N=4$ supersymmetries.
Extensions of this result were studied in [14], where it was
found that the bound on $N$ is not affected by the presence of
torsion, while for local supersymmetry the restriction $N\leq 4$
remains intact for conformally invariant theories. Without
conformal invariance there are certainly theories with $N>4$
[15,16], but those were never studied systematically.
Because three-dimensional supergravity has no conformal
invariance, one expects no restriction to $N\leq4$ (although the
$N=4$ models remain somewhat special as we shall see). On the other
hand, extended supergravities in four dimensions are known to be
restricted to $N\leq 8$ in view of the non-existence of
consistent interacting theories describing massless particles
with spin $s>2$ (we note, however, that this bound can
possibly be circumvented in certain theories which are not of the
conventional type [17]). The fact that three-dimensional
supergravities with even $N$ correspond to four-dimensional
theories with $N/2$ local supersymmetries, and can therefore
be constructed by dimensional reduction, suggests the
bound $N\leq 16$ in three dimensions. Indeed, a central
result of this paper is that extended theories {\it do} satisfy
this restriction, and this fact in turn constitutes an
alternative proof of the four-dimensional result. However,
the result now hinges on the geometric properties of target spaces
with restricted holonomy groups, a subject which has been studied
in considerable depth in the mathematical literature [18].

Because the geometrical arguments leading to these restrictions
are at the heart of this paper, we now briefly summarize them.
The general analysis of the Lagrangian and transformation rules
given in section~3 enables us to derive the constraints on the
Riemann curvature tensor, and hence on the holonomy group of the
target manifold, that are imposed by local supersymmetry
(see (4.19), which is the crucial formula). These conditions
become more and more restrictive with increasing $N$; for $N>4$,
they completely determine the target manifolds, whereas they are not
strong enough to determine them for $N\leq 4$. In particular,
for $N=1$, there are no restrictions at all, and the
target space may be an arbitrary Riemannian manifold. For $N=2$,
there is one complex structure, and the target manifold is K\"ahler.
For $N=3$ and 4, there are three almost-complex structures.
For $N=3$ the space is quaternionic, while for $N=4$ the target
space is locally a product of two quaternionic manifolds,
associated with inequivalent supermultiplets. Nonetheless,
there remains a great variety of possibilities for $N\leq 4$, as
the manifolds are not homogeneous in general.
For $N\geq 5$, on the other hand, (4.19) implies that the
holonomy group becomes ``too small" in a sense to be made
precise in section 5. We first show that all manifolds are
Einstein spaces and then we derive how $d$ (the dimension
of the target space) and $N$ are restricted: we find
that an arbitrary number of supermultiplets is permitted for
$N=5,6,8$, while only one is allowed for $N=9,10,12$ and 16.
For other values of $d$ and $N$ no theories can exist!
We can then appeal to a powerful mathematical
theorem [19] and use our knowledge of the holonomy group for
$N\geq5$ to conclude that all the corresponding target manifolds
must be symmetric spaces; their determination
is thus simply a matter of matching the allowed values of
$N$ and $d$ with a list of symmetric spaces.
In this way, we identify a unique symmetric space for each
of these values of $N$ and $d$. The isometry groups of
the target spaces corresponding to $N=5,6,8$ are equal to
$Sp(2,k)$, $SU(4,k)$ and $SO(8,k)$, respectively, where
$k$ is the number of supermultiplets. For $N=9,10,12$ and 16
the corresponding target spaces possess
the exceptional isometry groups $F_{4(-20)}$, $E_{6(-14)}$,
$E_{7(-5)}$ and $E_{8(+8)}$, respectively; remarkably, they can be
interpreted as projective spaces over the octonions [18].
In view of our previous remarks and the fact that
the maximally extended $N=16$ theory is invariant under the
``maximally extended"  exceptional Lie group $E_8$
[8,12], we are intrigued by the fact that the
apparent non-existence of massless particles of spin $s>2$ in
four dimensions may be related to the non-existence of
exceptional groups beyond $E_8$.

A characteristic feature of the non-linear sigma models with local
supersymmetry is that the target-space connection for the
fermions is no longer the usual Christoffel connection, but it
contains extra terms proportional to the almost-complex
structures associated with the extra supersymmetries (see
(3.27)). This aspect is crucial for the two-loop
finiteness of the dimensionally reduced
models, which hinges on the
fact that the contraction $R_{iklm}\,R_j^{\;klm}$
of the corresponding curvature tensors remains independent of the
modification of the fermionic connection [11].
{}From the formulae derived
later (in particular (3.30) and (4.11)) it follows that
this is always the case for $N=3$ and $N>4$. For $N=4$ the
situation is somewhat more subtle, as one is in general
dealing with two separate quaternionic subspaces. Nevertheless
upon using (3.30) and (4.38) one can easily establish that
property holds whenever the two subspaces are of equal
dimension. In contrast the $N=1, 2$ theories will in general
fail to be finite at one-loop.
We will not return to this topic here and
leave it to the reader to verify these results.

This paper is organized as follows. In section~2 we review
the construction of $D=3$ supermultiplets. Section~3
contains the results for the invariant Lagrangian and the
supersymmetry transformation rules. The geometrical implications
of the presence of $N$ local supersymmetries for the target
space are then worked out in section~4. In section~5 we
identify the possible target spaces for $N\geq5$. As those
are all symmetric we include a discussion of the conventional
formulation of extended supergravity coupled to
non-linear sigma models with homogeneous target
spaces and elucidate the connection
with the target-space approach used in the previous sections.
Some material relevant for the exceptional cosets is
relegated to an appendix.

\beginsection 2. Massless $D=3$ supermultiplets

Consider the extended supersymmetry algebra, with the
anti-commutation relation
$$
\big\{Q_\a^I\,,\bar Q_\b^J\big\} =- 2i\,\d^{IJ} \g^\m_{\a\b}\, P_\m
\,,\qquad (I,J= 1,\ldots, N) \eqno(2.1)
$$
where the $Q^I_\a$ are $N$ independent Majorana spinor charges
and $P_\m$ is the energy-momentum operator.
For states with light-like momentum, say in a frame where
$P^0=P^1=\o$ and $P^2=0$, (2.1) takes the following
form\footnote{${}^1$}{We use $\g_0=-i\s_2$, $\g_1= \s_1$,
$\g_2=\s_3$, with charge-conjugation matrix $C=\s_2$.}
$$
\big\{Q_\a^I\,, Q_\b^J\big\} = 2\o \,\d^{IJ} \big( {\bf 1} +
\s_3 \big)_{\a\b}. \eqno(2.2)
$$
In a positive-definite  Hilbert space of states, $Q^I_2$ must therefore
vanish and we are left with the real charges $Q^I_1$, which
generate an $N$-dimensional Clifford
algebra.\footnote{${}^2$}{Strictly speaking the charges are
hermitean; we insist on reality in view of field-theoretic
applications.}
In addition a fermion-number operator $\bf F$ must exist
satisfying ${\bf F}^2 =\bf 1$, which anti-commutes with the
supercharges $Q^I_\a$. Therefore massless supermultiplets are
representations of a
real $(N+1)$-dimensional Clifford algebra of positive signature.
In the basis where $\bf F$ is diagonal we
denote the bosonic indices by $A,B,\ldots = 1,\ldots ,d$ and
the fermionic indices by $\dot A, \dot B,\ldots = 1, \ldots,
d$. The supercharges then take the form of off-diagonal gamma
matrices
$$
\Gamma^I=
  \pmatrix{ 0 & \Gamma^I_{A  \dot D }   \cr
           \noalign{\vskip2mm}
           \Gamma^I_{{\dot B} C}   & 0  \cr }\,, \qquad
{\bf F} =   \pmatrix{ {\bf 1} & 0   \cr
              \noalign{\vskip3mm}
                0    & - {\bf 1}  \cr }\,.   \eqno(2.3)
$$
As one can always choose a basis where the gamma matrices are
symmetric, the two submatrices of $\G^I$ are each others
transpose; in terms of the upper-right $d\times d$ matrices
$\G^I_{A\dot B}$, which themselves have no special symmetry
properties, the defining relation of the Clifford algebra reads
$$
\G^I_{A\dot C}\,\G^J_{B\dot C} +\G^J_{A\dot
C}\,\G^I_{B\dot C} =2\d^{IJ}\d_{AB} \,. \eqno(2.4)
$$
The irreducible supermultiplets are listed in Table~1, together
with their centralizers [20].

For odd values of $N$ the supermultiplet is unique up to a
similarity transformation. For even values of $N$ the product of
the $N+1$ generators of the algebra,
$$
\tilde \G \equiv {\bf F}\,\G^1\, \cdots \G^N  \eqno(2.5)
$$
commutes with $\bf F$ and $\G^I$. For $N= 4$ mod~4 it satisfies
$\tilde \G{}^2 = \bf 1$, so that the Clifford algebra can be
decomposed into two simple ideals, associated with the projection
operators $\ft12 ({\bf 1}\pm \tilde \G)$. Inequivalent
irreducible representations of the Clifford algebra correspond to
one of these ideals and are characterized by $\tilde\G=\pm {\bf 1}$. For
$N = 2$ mod~4 we have $\tilde\G{}^2= -\bf 1$ and
the representation is again unique; it cannot be decomposed
into irreducible representations unless one introduces {\it
complex} projection operators. The existence of inequivalent
supermultiplets is a special feature of supersymmetry in low
space-time dimensions. In higher dimension the spinor
character of the supercharges ensures that inequivalent
supermultiplets have a different spin content, so that there is
no need for making a further distinction. From Table~1, we infer
that the multiplets with $N=3$ and
$N=4$ are the same; likewise
$N=5,6,7,8$ have identical multiplets (this result holds again
modulo 8, so that also $N=11,12$ have identical multiplets, and so on).
However, the situation is different in the case of local
supersymmetry, because the number of gravitini is not the same
for different values of $N$.

\topinsert
\baselineskip= 4 true mm
$$\vbox{\offinterlineskip
\hrule
\halign{&\vrule# & \strut\quad\hfil#\hfil\quad \cr
height2pt&\omit  &&\omit&&\omit&&\omit & \cr
&$N$ && $d_N$ && centralizer    & \cr
height2pt&\omit  &&\omit&&\omit & \cr
 \noalign{\hrule}
height2pt&\omit  &&\omit&&\omit&&\omit & \cr
&1   && 1  && $\bf R$    & \cr
&2   && 2  && $\bf C$    & \cr
&3   && 4  && $\bf H$    & \cr
&4   && 4  && $\bf H$    & \cr
&5   && 8  && $\bf H$    & \cr
&6   && 8  && $\bf C$    & \cr
&7   && 8  && $\bf R$    & \cr
&8   && 8  && $\bf R$    & \cr
&$n+8$ && $16 d_n$ && as for $n$  & \cr
height2pt&\omit &&\omit&&\omit &\cr}
\hrule}
$$
\narrower\narrower\noindent
Table~1. Irreducible massless supermultiplets with $d_N$ the
number of bosonic states. The centralizer
contains the operators that commute with the
supercharges {\it and} with fermion number, which constitute a
division algebra.

\endinsert

Observe that fermions and bosons in an irreducible multiplet
transform according to irreducible spinor representations of
$SO(N)$. Here we recall the well-known result that the spinor
representations of $SO(N)$ are real for $N=1,7,8$ mod 8, complex
for $N=2,6$ mod 8 and pseudo-real for $N=3,4,5$ mod 8 (see e.g.
[21]).
{}From Table~1 it is obvious that these cases correspond
to the centralizers $\bf R$, $\bf C$ and $\bf H$, respectively.
For $N=2,\ldots,6$ mod 8, the centralizer contains (at least)
the identity
and a real antisymmetric matrix $e$ with $e^2=-\bf1$,
acting within the bosonic and fermionic subspaces.
Clearly, $e$ can be traded for the imaginary unit
$i$ by complexifying the representation. By use of the complex
projection operators $\ft12 ({\bf 1}\pm ie)$ the
real $d$-dimensional $SO(N)$ representations become
$d/2$-dimensional {\it complex} representations, and the matrices
$\G^I_{A\dot A}$ can be replaced by complex $d/2 \times
d/2$ matrices. This observation will be important for the
derivation of the completeness relations and Fierz rearrangement
formulas used in the appendix. For $N=3,4,5$ mod 8, there are two
additional complex structures that anticommute with $e$. Either
one of them can be used to show that the representation is
actually pseudo-real.

In the remainder of this section we present the explicit
construction of the supercharges for $N=1,2,4,8$~mod~8, to
facilitate the discussion in the subsequent sections (for further
explicit details, see [22]). The representations for intermediate
values of $N$ have the same dimensionality as one of the
$N=1,2,4,8$~mod~8 representations and can conveniently be
studied by embedding them in the higher-$N$ representation; the
centralizer can be explicitly constructed from
the centralizer of the higher-$N$ representation, possibly
extended with some of the extra gamma matrices.

We start by defining a basis of the $2\times 2$ real matrices,
consisting of the identity $\bf 1$, $\s_1$, $\s_3$ and
$\varepsilon\equiv -i\s_2$. Hence we have
$$
\varepsilon = \s_1\,\s_3\, \qquad \var\, \s_{1} = -\s_{3}\,,
\qquad  \var \,\s_{3}= \s_{1}\,. \eqno(2.6)
$$

For $\underline{N=1}$ we choose ($d_1=1$)
$$
{\bf F}(2) = \s_3\,, \qquad \G^1(2) = \s_1\,,     \eqno(2.7)
$$
where the number in parentheses indicates the dimension of the
matrix. Hence, for $N=1$ one has $\G^1_{A\dot A}=1$. We note the
properties
$$
\var^2=-{\bf 1}\,,\quad \{\var,\G^1\} =\{\var,{\bf F}\} =0\,.
\eqno(2.8)
$$

For $\underline{N=2}$ a representation of the Clifford algebra is
constructed by taking direct products of $2\times 2$ matrices
times the previous lower-dimensional algebra (so that $d_2=2$):
$$
\eqalign{ {\bf F}(4)&= \s_3\otimes {\bf 1}(2)  \,,\cr
          \G^1(4)   &= \s_1\otimes \G^1(2)  \,,   \cr
          \G^2(4)   &= \s_1\otimes {\bf F}(2) \,, \cr} \qquad
\hbox{with}\qquad
\G^{12} = {\bf 1}\otimes \var\,. \eqno(2.9)
$$
so that
$$
\G^1_{A\dot A} = \pmatrix{0&1\cr 1&0\cr}\,, \qquad \G^2_{A\dot A}
= \pmatrix{1&0\cr 0&-1\cr}\,. \eqno(2.10)
$$
In addition we note the existence of the following three complex
structures
$$
\eqalign{ e_1(4) &= \s_3\otimes \var  \,,\cr
          e_2(4) &=-\var\otimes {\bf 1}(2)  \,,   \cr
          e_3(4) &= \s_1\otimes \var \,, \cr} \qquad
\hbox{satisfying} \qquad e_i\,e_j = - \d_{ij}\,{\bf 1}
+\e_{ijk}\,e_k. \eqno(2.11)
$$
Note that ${\bf F}\,\G^1\,\G^2=e_1$, and
$$\eqalignno{
& [e_1, \G^1] = [e_1,\G^2] = [e_1,{\bf F}] = 0\,,\cr
& \{e_2, \G^1\} = \{e_2,\G^2\} = \{e_2,{\bf F}\} = 0\,,\cr
& \{e_3, \G^1\} = \{e_3,\G^2\} = \{e_3,{\bf F}\} = 0\,. &(2.12)\cr}
$$
The centralizer of the Clifford algebra is based on $e_0\equiv
\bf 1$ and $e_1$, so that the associated symmetry group is $U(1)$.
Note, however, that in the bosonic or the fermionic subspace
$e_1$ and $\G^{12}$ are degenerate.

Fur future use note the identities
$$
e_1\,\G^1 =\G^2\,{\bf F}\,,\qquad
e_1\,\G^2 ={\bf F} \,\G^1\,,\qquad
e_1\,{\bf F} =\G^{12}\,.   \eqno(2.13)
$$

For $\underline{N=4}$ we take again direct products of $2\times
2$ matrices times the matrices of the previous algebra (so that
$d_4=4$):
$$
{\bf F}(8)= \s_3\otimes {\bf 1}(4)  \,,\qquad
\eqalign{ \G^1(8)   &= \s_1\otimes \G^1(4)  \,,   \cr
          \G^2(8)   &= \s_1\otimes \G^2(4)  \,, \cr
          \G^3(8)   &= \s_1\otimes {\bf F}(4)  \,,   \cr}
\qquad  \G^4(8)  = \var\otimes e_1(4)  \,, \eqno(2.14)
$$
with the complex structures
$$
\eqalign{ e_1(8) &= {\bf 1} \otimes e_1(4)  \,,\cr
          e_2(8) &= \s_3 \otimes e_2(4)     \,,   \cr
          e_3(8) &= \s_3\otimes e_3(4)      \,, \cr} \qquad
\hbox{satisfying} \qquad e_i\,e_j = - \d_{ij}\,{\bf 1}
+\e_{ijk}\,e_k. \eqno(2.15)
$$
Observe that ${\bf F}\,\G^1\, \G^2\, \G^3\, \G^4 =-\bf 1$. As
explained previously there
are two inequivalent representations. A second one is, for
instance, found by changing the sign of $\G^1$, $\G^2$, $\G^3$.

This time all $e_i$ commute with $\G^I$ and $\bf F$,
$$
[e_i, \G^I] = [e_i,{\bf F}] = 0\,.    \eqno(2.16)
$$
so that the centralizer of the algebra consists of $e_0\equiv \bf 1$
and $e_i$ associated with the group $SU(2)$.

The $SO(4)$ generators are
$$
\eqalign{
 & \G^{12}  = {\bf 1}\otimes \G^{12} \,,   \cr
 & \G^{23}  = {\bf 1}\otimes \G^2  \,{\bf F}
            = {\bf 1}\otimes e_1\,\G^1 \,, \cr
 & \G^{31}  = {\bf 1}\otimes {\bf F}\, \G^1
            = {\bf 1}\otimes e_1\,\G^2 \,,\cr}  \qquad
\eqalign{
 & \G^{34}  = \s_3\otimes {\bf F} e_1 =  \s_3 \otimes \G^{12} \,, \cr
 & \G^{14}  = \s_3\otimes e_1\,\G^1  \,,   \cr
 & \G^{24}  = \s_3\otimes e_1\,\G^2  \,,   \cr } \eqno(2.17)
$$
where we made use of the identities derived previously for $N=2$.
This shows that
$$
{\bf F}\,\G^{IJ} = \ft12  \e^{IJKL}\, \G^{KL}\,.  \eqno(2.18)
$$
Therefore the $SO(4)$ group factors into two $SO(3)$ groups, one
acting on the bosons (the selfdual component) and one
on the fermions (the anti-selfdual component).
This feature will play an important role in
the discussion of $N=4$ theories in sections~4 and 5.

For $\underline{N=8}$, we have $d_8=8$ from Table 1. The gamma matrices
are then explicitly given by
$$
{\bf F}(16)  = \s_3\otimes {\bf 1}(8)  \,,\qquad
\eqalign{ \G^1(16)   &= \s_1\otimes \G^1(8)   \,,   \cr
          \G^2(16)   &= \s_1\otimes \G^2(8)  \,, \cr
          \G^3(16)   &= \s_1\otimes \G^3(8)  \,,   \cr
          \G^4(16)   &= \s_1\otimes \G^4(8)  \,, \cr
          \G^5(16)   &= \s_1\otimes {\bf F}(8) \,,  \cr}
\qquad
\eqalign{ \G^6(16)   &= \var\otimes e_1(8)  \,, \cr
          \G^7(16)   &= \var\otimes e_2(8)  \,,   \cr
          \G^8(16)   &= \var\otimes e_3(8)  \,. \cr}   \eqno(2.19)
$$
Just as for $N=4$ this representation is not unique; a second
inequivalent representation exists, and may, for instance, be
obtained by changing the sign of $\G^6$, $\G^7$ and $\G^8$.

For $\underline{N>8}$ the pattern repeats itself; for $N=n+8$, the
dimensionality of the gamma matrices equals $16\,d_n$ and we put
($n\leq8$)
$$
{\bf F} = {\bf F}(2d_n)\otimes {\bf 1}(16)\,,\qquad
\eqalign{\G^I&=\G^1(2d_n)\otimes \G^I(16)\,,\cr
         \G^9&=\G^1(2d_n) \otimes {\bf F}(16)\,,\cr}\qquad
\G^{8+a}= \G^a(2d_n)\otimes {\bf 1}(16)\,, \eqno(2.20)
$$
where $I=1,..,8$ and $a=2,...,n$, while $\G^1(2d_n)$ and
$\G^a(2d_n)$ are the $(2d_n\times2d_n)$ gamma matrices corresponding to the
irreducible representation of the $n$-dimensional Clifford
algebra. The centralizer is of the form $Z(2d_n)\otimes {\bf 1}(16)$,
where $Z(2d_n)$ is the centralizer of the $n$-dimensional Clifford
algebra.

Finally, let us add that for {\it reducible} representations,
the centralizer generates the group $SO(k)$, $U(k)$ or $Sp(k)$,
depending on whether the centralizer for an irreducible
representation corresponds to $\bf R$, $\bf C$ or $\bf H$,
respectively. Here $k$ denotes the number of irreducible
representations. The case of $N= 4$ mod 4 is again exceptional
because one is dealing with inequivalent representations [23].
For $k_1$ and $k_2$ inequivalent
representations, the corresponding groups are $SO(k_1)\otimes
SO(k_2)$ (for $N=8$  mod 8) and $Sp(k_1)\otimes Sp(k_2)$ (for
$N=4$ mod 8).

\beginsection 3. Lagrangian and transformation rules

In this section we present the full Lagrangian and transformation
rules for a non-linear sigma model coupled to $N$-extended
supergravity. Let us first introduce the separate Lagrangians for
pure supergravity and the non-linear sigma model. The supergravity
Lagrangian can be written as follows\footnote{${}^3$}{We use the
Pauli-K\"all\'en metric with
$\g_{[a}\g_b\g_{c]} = i\e_{abc}$, $\g_{ab}\equiv
\g_{[a}\g_{b]} = i\e_{abc}\,\g^c$. Readers who prefer the $(-,+,
+)$ metric multiply Dirac conjugate spinors and $\e_{abc}$ by $i$, and
$\e^{abc}$ by $-i$.}
$$
\lagr_{\rm s.g.} = -\ft12 i\e^{\m\n\r}\,\Big\{ e_\m^{\;a} \,
R_{\n\r a}(\o) + \bar\p^I_\m D_\n(\o) \p^I_\r\Big\}  \,,
\eqno(3.1)
$$
with the $SO(2,1)$ covariant derivative acting on a spinor as
$$
D_\m(\o)\,\p = \big(\pa_\m + \ft12 \o_\m^a\,\g_a \big) \p \,.
\eqno(3.2)
$$
The spin-connection field $\o^a_\m$ will be regarded as an
independent field (first-order formalism). Its field equation
implies that the supercovariant torsion tensor vanishes, i.e.,
$$
D_{[\m}(\o)\, e^a_{\n]}  -\ft14 \bar \p{}^I_\m \g^a\p^I_\n= 0 \,,
\eqno(3.3)
$$
where
$$
D_{\m}(\o)\, e^a_{\n} = \pa_{\m}e^a_{\n}+ i \e^{abc}
\o_{\m\,b}\,e_{\n\,c} \,. \eqno(3.4)
$$
{}From (3.4) one determines the spin connection; substituting
the result into the field strength
$$
R^a_{\m\n}(\o) = \pa_\m\o^a_\n -\pa_\n\o^a_\m + i \e^{abc}\,
\o_{\m\,b}\,\o_{\n\,c}  \eqno(3.5)
$$
yields the Riemann tensor (up to
gravitino-dependent terms). The Lagrangian (3.1) is locally
supersymmetric under $N$ independent supersymmetries. There is no
restriction on the number of independent local supersymmetries
and the theory is topological [3].

The rigidly supersymmetric non-linear sigma model is described
by the Lagrangian
$$
{\cal L}_{\rm matter} =  - \ft12
g_{ij}(\phi)\Big\{\partial_\mu\phi^i\,\partial^\mu\phi^j +
 \bar\chi^i  \rlap /\! D(\Gnul)\,\chi ^j \Big\}
+ \lagr_{\x^4} \,,   \eqno(3.6)
$$
where the target-space connection $\Gnul$ equals the Christoffel
symbol and the covariant derivative is defined by (for arbitrary
connection $\G$)
$$
D_\m(\G)\,\x^i \equiv \pa_\m \x^i + \G^i_{jk}(\f)\, \pa_\m\f^j\,\x^k\,
.\eqno(3.7)
$$
We denote the dimension of the target space by $d$, so that $i,j,
\ldots =1,\ldots , d$.
The $\x^4$-terms are proportional to the Riemann tensor of the
sigma-model target space,
$$
\lagr_{\x^4} = -\ft1{24} R_{ijkl}(\phi) \;\bar\chi^i\gamma_a \chi^j
\;\bar\chi^k\gamma^a \chi^l \,.  \eqno(3.8)
$$
Ignoring the extra space-time coordinate, the Lagrangian (3.6) is
identical to the one in two dimensions; the $\x^4$-terms can be
rewritten in a form where we sum over only two independent gamma
matrices, by using the cyclicity of the
Riemann tensor.

The non-linear sigma model have $N=1$, 2 or 4 independent rigid
supersymmetries. The extra supersymmetries are associated with
complex structures $\cstr Pij$, labeled by $P=2,\ldots N$, which
are hermitean,
$$
g_{ij}\,\cstr Pjk +g_{kj}\,\cstr Pji  = 0\,, \eqno(3.9)$$
and satisfy the Clifford property
$$
\cstr Pik \,\cstr Qkj +  \cstr Qik\, \cstr Pkj = -2\,
\delta_{PQ}\,\d^i_j \,. \eqno(3.10)
$$
Furthermore they are covariantly constant (with
respect to the Christoffel connection),
$$
D_i(\Gnul)\cstr Pjk \equiv \partial_i\cstr Pjk +
\Gnul_{il}^{j}\,\cstr Plk  - \Gnul_{ik}^{l}\,\cstr Pjl  =0 \,.
\eqno(3.11)
$$
The upper limit on $N$ arises because the holonomy group commutes
with the complex structures. Therefore this group must either act
reducibly in target space, in which case the target space becomes
reducible (i.e. it decomposes into separate spaces),
or, by Schur's lemma (see e.g. [24]), the complex structures
must generate a
division algebra; the largest such algebra is the quaternionic
one with three complex structures, corresponding to $N=4$
[13]. Alternatively, one may make use of
the fact that these models are invariant under $SO(N)$ rotations
on the fermions (for $N=4$ one has only $SO(3)$). Combining these
transformations with supersymmetry proves that the theory must be
invariant under non-uniform translations of space-time
coordinates as soon as $N>4$, which implies that the target space
is reducible [14].

So far we have put Newton's constant to unity, but in what
follows we want to be a little more explicit about the
dimension of the various quantities. It is convenient to choose
all boson fields dimensionless, with
the exception of the spin connection which has dimension
$[1]$ (in mass units); the fermion fields have dimension $[1/2]$
and the supersymmetry transformation parameter dimension
$[-1/2]$. In this way none of the transformation rules will contain
dimensional parameters, whereas the Lagrangian contains just an
overall constant $1/\k$, where $\k$ has dimension $[-1]$.
Hence we write
$$
\lagr = {1\over\k} \Big\{\lagr_{\rm s.g.}+\lagr_{\rm kin} +\lagr_{\rm
N} +\lagr_{\x^4}  \Big\} \eqno(3.12)
$$
Here $\lagr_{\rm s.g.}$ is the supergravity Lagrangian, modified
by extra matter-dependent connection terms (here and henceforth
we decompose the indices $I$ into $I=1$ and $I=P= 2,\ldots ,N$;
the gravitino field and corresponding supersymmetry parameter
with $I=1$ are denoted by $\p_\m$ and $\e$, respectively),
$$
\lagr_{\rm s.g.} = -\ft12 i\e^{\m\n\r}\,\Big\{ e_\m^{\;a} \,
R_{\n\r a}(\o) + \bar\p_\m D_\n(\o,Q) \p_\r +\bar\p^P_\m D_\n(\o,
Q) \p^P_\r\Big\}  \,, \eqno(3.13)
$$
where
$$\eqalignno{
D_\m(\o,Q) \,\p_\n &= D_\m(\o) \,\p_\n  -\pa_\m\f^i\, Q^P_i(\f) \,
\p^P_\n   \,,\cr
D_\m(\o,Q)\, \p^P_\n &= D_\m(\o)\, \p^P_\n  +\pa_\m\f_i\, \big[
Q^{PQ}_i(\f) \, \p^Q_\n +  Q^P_i(\f)\,\p_\n\big]\,.  &(3.14)\cr}
$$
Clearly $Q_i^P$ and $Q^{PQ}_i$ can be combined into an $SO(N)$
target-space connection $Q_i^{IJ}$.

The $\lagr_{\rm kin}$ refers to the properly covariantized
kinetic terms of the non-linear sigma model,
$$
\lagr_{\rm kin} = - \ft12 e\,
g_{ij}(\phi)\Big\{g^{\m\n}\partial_\mu\phi^i\,\partial_\n\phi^j +
 \bar\chi^i  \rlap /\! D(\o,\G)\,\chi ^j \Big\}\,, \eqno(3.15)
$$
where the connection $\G$ is no longer the Christoffel
connection but may contain extra terms. As only the
anti-symmetric part of $\G$ appears in (3.15), we may assume
without loss of generality that the metric postulate remains
satisfied,
$$
D_i(\G)\,g_{jk} = 0\,. \eqno(3.16)
$$
The torsion now receives contributions from the
spinor fields $\x^i$, so that (3.3) changes into
$$
D_{[\m}(\o)\, e^a_{\n]} -\ft14 \bar \p{}^I_\m \g^a\p^I_\n -\ft18i
e\,\e_{\m\n\r}\, e^{\r a} \,g_{ij} \bar\x{}^i\x^j = 0 \,
.\eqno(3.17)
$$

Just as in the case of rigid supersymmetry, the extra
supersymmetries are associated with tensors $\cstr Pij$. However,
in the context of local supersymmetry these tensors are usually
not complex structures, but only almost-complex structures (for
definitions, see e.g. [25]);
indeed, as we shall see later, their Nijenhuis
tensors do not vanish in general. The almost-complex structures
appear in the Lagrangian $\lagr_{\rm N}$, which refers to the
Noether terms with certain higher-order modifications to ensure
the supercovariance of the $\x^i$ field equation,
$$\eqalignno{
{\cal L}_{\rm N} & =\ft14 e\,g_{ij}\bar\chi^i\gamma^\mu\,
\big(\pa\slash\f^k +\hat\pa\slash\phi^k\big) \big(\delta^j_k\,\psi_{\mu} -
\cstr Pjk \,\psi^P_{\mu}\bigr )         &(3.18) \cr
& =  \ft12 e\,g_{ij}\bar\chi^i\gamma^\mu\,
\pa\slash\f^k \big(\delta^j_k\,\psi_{\mu} -
\cstr Pjk \,\psi^P_{\mu}\bigr ) \cr
&\quad +\ft1{16}
e\, g_{ij}\,\bar\chi^i\chi^j\,
\big(\bar\psi_{\nu}\g^\m\g^\n \p_\m  + \bar\psi^P_{\nu}\gamma^\mu\gamma
^\nu \psi_\mu^P\big) \cr
&\quad +\ft1{16} e\,\bar\chi^i\gamma_\rho\chi^j\,\big[ \big(f_{[P}\,
f_{Q]}\big)_{ij}  \bar\psi
_{\nu}^P \g^\r\g^\m\g^\n\psi_{\mu}^Q + f_{P\,ij} \,
\bar\p^P_\n (\g^\r\g^\m\g^\n +\g^\m\g^\n\g^\r) \psi_\mu
\big]\,,  \cr}
$$
where we used that the supercovariant derivative of $\f^i$ is
equal to
$$
\hat\pa_\m\f^i = \pa_\m\f^i -\ft12 \big (\delta^i_j \,\bar\p_\m
+ \cstr Pij \, \bar\p_\m^P \bigr )\chi^j \,.\eqno(3.19)
$$
Also the $\x^4$-terms are modified due to the local supersymmetry,
and we find
$$
\lagr_{\x^4} = \ft1{16}e\,\big(g_{ij}\, \bar\x{}^i\x^j\big)^2
-\ft1{24}e\, R_{ijkl} \;\bar\chi^i\gamma_a \chi^j
\;\bar\chi^k\gamma^a \chi^l \,.  \eqno(3.20)
$$
The supersymmetry transformation rules are
$$\eqalignno{
\delta e_\mu^a & = \ft12 \bar\epsilon \gamma^a\psi_{\mu} +
   \ft12\bar\epsilon^P \gamma^a\psi_{\mu}^P\,, &(3.21)\cr
\delta\psi_{\mu} & = D_\m(\o,Q)\, \epsilon  +\d\f^i\,Q^P_i\,
   \p^P_\m -\ft18 \bar\x^i\g^\n\x^j\,f_{P\,ij} \,\g_{\m\n}\,\e^P
 \, ,&(3.22)\cr
\delta\psi_{\mu}^{P} & =  D_\m(\o,Q) \,\epsilon^P
   -\d\f^i\,\big[ Q^P_i\,\p_\m +Q^{PQ}_i\, \p^Q_\m\big]   \cr
&\quad +\ft18 \bar\x^i\g^\n\x^j\,\big[\big(f_{[P}f_{Q]}\big)_{ij}
  \g_{\m\n}\,\e^Q + f_{P\,ij} \,\g_{\m\n}\,\e\big]\,  ,&(3.23)\cr
\delta\phi^i & = \ft12\big (\delta^i_j \,\bar\epsilon
+ \cstr Pij\, \bar\epsilon^P \bigr )\chi^j\, , &(3.24)\cr
  \delta\chi^i &=\ft12 \hat{\rlap/\partial}
\phi^j\bigl(\epsilon\, \delta^i_j - \cstr Pij\,\epsilon^P
\bigr ) - \Gamma_{jk}^{i} \,\delta\phi^j
\chi^k  \,.&(3.25) \cr}
$$

Let us now briefly comment on the derivation of these results.
One starts with the sum of (3.1) and (3.15) and follows the same
strategy as in [14] by introducing an as yet undetermined
connection $\G$ into the Lagrangian and transformation rules.
The first variations are standard and quickly reveal the need for
the Noether terms. At that point one has variations proportional
to $\pa\f\,\pa\f\, \x\,\e$ and $\pa\f\,\pa\f\, \p\,\e$. The former can be
cancelled by introducing the $Q$-dependent terms in the gravitino
transformation rules, which at the same time requires one to add
corresponding $Q\,\p\,\p$ terms to the action. This restricts
the form of $Q_i$ to $SO(N)$ target space connections (cf. (3.14)), and
leads in turn to new $\pa\f\,\pa\f\, \p\,\e$ variations. Both the
$\pa\f\,\pa\f\, \x\,\e$ and $\pa\f\,\pa\f\, \p\,\e$ variations
vanish provided the $SO(N)$ curvatures satisfy the condition
$$\eqalignno{
R_{ij}^P(Q)&\equiv \pa_{i}Q_{j}^P +Q_{i}^{PQ}\,Q_{j}^Q
-(i\leftrightarrow j) = -\ft12 f_{P\,ij}\,, \cr
R_{ij}^{PQ}(Q)&\equiv \pa_{i}Q_{j}^{PQ} +Q_{i}^{PR}\,
Q_{j}^{RQ} -Q_{i}^{P}\,Q_{j}^Q -(i\leftrightarrow j)= \ft12 \big(f_{[P}\,
f_{Q]}\big){}_{ij} \,,
&(3.26) \cr}
$$
the connection $\G$ is given by
$$
\G^j_{ik} = \Gnul^j_{ik} - Q^P_i\,\cstr Pjk \,,      \eqno(3.27)
$$
and the almost-complex structures are covariantly constant in the
following sense,
$$
D_k(\Gnul) \,f_{P\,ij} + Q^{PQ}_k\, f_{Q\,ij} + Q^Q_k\,\big(f_{[P}\,
f_{Q]}\big){}_{ij} =0 .    \eqno(3.28)
$$
The latter result ensures that the Bianchi identities of the $SO(N)$
curvatures remain consistent with the constraints (3.26). It also
allows the evaluation of the Nijenhuis tensors (no summation over
$P$ implied)
$$
N^{\,k}_{P\,ij} = \cstr Pli\, f^{\,k}_{P [l;j]} - \bigl ( \;
i\leftrightarrow j\;\bigr ) \,,   \eqno(3.29)
$$
which satisfy $N^{\;j}_{\!P\,ji}=0$, but vanish only for $N=2$ where
the complex structure is covariantly constant with respect to
the Christoffel connection. Let us also note that the curvature
associated with the connection (3.27) is equal to
$$
R_{ijkl}(\G) = R_{ijkl} -\ft12 f^P_{ij}\,f^P_{kl}\,, \eqno(3.30)
$$
where we used (3.28).

At this point all variations of the Lagrangian linear in the
spinor fields vanish. Subsequently one concentrates on the terms
proportional to three spinors with a derivative acting on one of
them. This then requires one to introduce the $\d\f\,Q\,\p\,\e$ and
the $\x^2\,\e$ variations in (3.22-23) and the $\p\,\x\,\e$
variations contained in the supercovariant derivative in (3.25).
The gravitino fields in the Lagrangian
and transformation rules are restricted
by supercovariance arguments; therefore, in view of
dimensional arguments, the only extra variations that one
expects are possible $\x^2\,\e$ terms in (3.25). However, it turns
out that those are not needed and one determines directly the
$\x^4$ terms in the action (cf. (3.20)) by making use of the
integrability conditions that are derived directly from
(3.28) and (3.26). We refrain from giving these conditions here,
as they will be discussed in the next section (cf. (4.4)).
By virtue of the integrability conditions also the remaining
variations, all cubic and quintic in the spinor fields, cancel
after tedious but straightforward calculation!

\beginsection 4. Target space geometry

In this section we study the implications of local supersymmetry
on the target-space geometry. The most obvious restriction
concerns the dimension of the target space. Locally it must be
decomposable into a number of supermultiplets. Therefore we must
have $d=k \,d_N$, where $k$ is an integer denoting the number of
irreducible supermultiplets and $d_N$ is the number of bosonic
states of an irreducible supermultiplet listed in Table~1. For
$N=1$,~2 the remaining implications are rather straightforward.
When $N=1$ the target space is a Riemannian manifold of arbitrary
dimension (as $d_1=1$) and
no special properties are required, while for $N=2$ we are
dealing with a K\"ahler space, as there is a complex
structure that is covariantly constant with respect to the
Christoffel connection (cf. (3.28)). Obviously such a space must be
of even dimension.  It then follows that the Ricci
tensor is related to the first Chern class.

The analysis for $N>2$ is more involved. It is convenient to
adopt a manifest $SO(N)$ notation. First introduce the
anti-symmetric tensors $f^{IJ}_{ij}$ (we freely raise and lower
$SO(N)$ indices),
$$
f^{PQ}= f^{[P}\,f^{Q]} \,, \qquad f^{1P} =\pm f^P\,,\eqno(4.1)
$$
and the $SO(N)$ target-space connections $Q^{IJ}_i$, consisting
of $Q^{PQ}_i$ and
$$
Q^{1P}_i = \mp Q^P_i\,. \eqno(4.2)
$$
With these definitions (3.26) and (3.28) can be written as
$$\eqalignno{
R_{ij}^{IJ}(Q)&\equiv \pa_{i}Q_{j}^{IJ} -\pa_{j}Q_{i}^{IJ}+
2\,Q_{i}^{K[I}\,Q_{j}^{J]K} = \ft12 f^{IJ}_{ij} \,,  \cr
D_if^{IJ}_{jk} &\equiv D_i(\Gnul) \,f^{IJ}_{jk} +2\, Q^{K[I}_i\,
f^{J]K}_{jk} =0\,.&(4.3) \cr}
$$
They lead to the integrability condition
$$
R_{ijmk}\,f^{IJ\,m}{}_{\!l} -R_{ijml}\,
f^{IJ\,m}{}_{\!k} = -f^{K[I}_{ij}  \,f^{J]K}_{kl} \,,\eqno(4.4)
$$
which, as pointed out in the previous section, was required for
the cancellation of the supersymmetry variations of the action
that are cubic and quintic in the spinor fields.

Obviously the tensors $f_{IJ}$ act as generators of $SO(N)$ in target
space,
$$
f_{IJ}\,f_{KL} -f_{KL}\,f_{IJ}  = 4\,\d_{K[I}\,f_{J]L} -
4\,\d_{L[I}\,f_{J]K}  \,.\eqno(4.5)
$$
In addition they satisfy
$$\eqalignno{
\big(f_{IJ}\big)^2 &=-{\bf 1} \,,\quad \hbox{($I$ and $J$ fixed)}\cr
f_{IK}\,f_{KJ} &= (N-1)\,\d_{IJ} -(N-2)\,f_{IJ}\,, \cr
f_{IJ\,ij}\,f_{KL}{}^{\!ij}  &= 2d \,\d_{I[K}\,\d_{L]J} \mp \d_{N,
4}\, \e_{IJKL} \, J^k_{\;k}\,. &(4.6)  \cr}
$$
The tensor $J$ is defined by
$$
\big(f_{[P_1}\, \cdots f_{P_{N-1}]}\big){}^i_{\;j} = J^i_{\;j}\;
\e_{P_1\cdots P_{N-1}} \,.\eqno(4.7)
$$
For {\it even} values of
$N$ it satisfies the following properties,
$$
J^i_{\;k}\,\cstr Pkj = \cstr Pik\,J^k_{\;j} \,,\qquad
D_i(\Gnul)J^j_{\;k} = 0\,,\qquad J^2 = (-)^{N/2}\,{\bf 1} \,,
\qquad J_{ij} = (-)^{N/2}\, J_{ji} \,, \eqno(4.8)
$$
and must be traceless, unless $N=4$. For $N=4$ one
derives
$$
f_P\,f_Q = -\d_{PQ} \,{\bf 1} - \e_{PQR}\,J\, f_R \,. \eqno(4.9)
$$
Hence $J^k_{\;k}$ is the trace of the product of the three
almost-complex structures, which is constant so that it may be
evaluated at any  point in target space. As $J$ is symmetric for
$N=4$ and its  square is equal to the unit matrix (cf. (4.8)), we
find
$$
J^k_{\;k} = d_+ -d_- \,,\eqno(4.10)
$$
where $d_\pm$ are the dimensions of the subspaces for which the
eigenvalue of $J$ is equal to $\pm 1$. More generally, for
$N=4$~mod~4, the subspaces with $J=\pm 1$ correspond to the
inequivalent supermultiplets discussed in section~2.

Let us now proceed for a general value of $N>2$. First we note
that for $N=3$ the tensors $f^{IJ}$ define precisely three
almost-complex structures, which are covariantly constant with
respect to a non-trivial $SO(3)\sim Sp(1)$ connection (cf.
(4.3)). Hence the target space must be quaternionic for $N=3$.
Leaving the special case of $N=4$ until the end of this section,
we now continue as generally as possible for $N>2$.
Contracting (4.4) with $f^{MN}_{kl}$ and making use of (4.6)
gives
$$
R_{ijkl}\,f_{IJ}{}^{\!kl} = \ft14d\,f_{IJ\,ij} \,, \eqno(4.11)
$$
while contracting (4.4) with $g^{jl}$, using the cyclicity of the
Riemann tensor and the above result (4.11), yields
$$
R_{ij} \equiv R_{ikjl}\,g^{kl} = c\, g_{ij} \,,\eqno(4.12)
$$
where
$$
c= N -2 +\ft18 d > 0\,.   \eqno(4.13)
$$
Hence we are dealing with an Einstein
space\footnote{${}^4$}{For $N=3$ this is in accord with the
fact that quaternionic spaces of dimension higher than four are
always Einstein [26]. In the case at hand, the result also
holds true for a four-dimensional target space. Our conventions
here are such that positive curvature ($c>0$) corresponds to
non-compact manifolds; this convention is opposite to the
one commonly adopted in the mathematical literature.}.

Now decompose the Riemann curvature as
$R_{ijkl} = \hat R_{ijkl} +\ft18 f^{IJ}_{ij}\,f^{IJ}_{kl}$,
so that (4.4) reads
$$
\hat R_{ijmk}\,f^{IJ\,m}{}_{\!l} - \hat R_{ijml}\,
f^{IJ\,m}{}_{\!k} = 0 \,.\eqno(4.14)
$$
This motivates us to introduce the set of independent
antisymmetric tensors $h_{ij}^\a(\f)$, labelled by indices $\a$
defined by the requirement that they commute with the $SO(N)$
generators,
$$
h^\a_{ik}\,f^{IJ\,k}{}_{\!j} -h^\a_{jk}\,f^{IJ\,k}{}_{\!i} =0 \,.
\eqno(4.15)
$$
For the moment we
restrict ourselves to a given point in target space, but the fact
that the $SO(N)$ generators are realized everywhere on the
manifold (in the spinor representation), implies that the number
of independent tensors $h^\a$ and their associated Lie-bracket
structure is the same everywhere.
Obviously the $h^\a$ generate the subgroup $H'$ of $SO(d)$ that
commutes with $SO(N)$; it will play an important role in what
follows. Because of Schur's lemma, $H'$ must be one of the groups
$SO(k_1)\otimes SO(k_2)$, $U(k_1)\otimes U(k_2)$ or
$Sp(k_1)\otimes Sp(k_2)$, where $k_1$ and $k_2$ denote the number
of inequivalent $SO(N)$ representations of the target space, and
we have $k=k_1+k_2$, as every irreducible supermultiplet contains
precisely one irreducible $SO(N)$ multiplet of scalar fields. The
nature or the group is determined by the centralizer of the
$SO(N)$ representation and can be read off from
Table~1; for $N=7,8,9$~mod~8 the group is orthogonal, for
$N=2,6$~mod~8, it is unitary, and for $N=3,4,5$~mod~8 it is
symplectic. For odd $N$ the spinor representation is unique, so
that one has $k_1=k$ and $k_2=0$. The structure constants of $H'$,
which may at this point depend on the target-space coordinates,
are defined by
$$
h^\a h^\b-h^\b h^\a = f^{\a\b}_\g \, h^\g\,. \eqno(4.16)
$$

{}From the arguments given above, as well as from more general
considerations, it follows that the compact group $H'$ factorizes
into a direct product of an Abelian
group and a number of simple groups. In what follows these factor
groups will generically be denoted by $H''$. By a suitable
redefinition we ensure that an index $\a$ refers exclusively to
one of these factor groups. Without loss of generality it is
possible to impose the normalization condition
$$
h^\a_{ij}\,h^{\b\,ij} =2 \,d_N \,\d^{\a\b} \,.\eqno(4.17)
$$
With this normalization it follows that $\d^{\a\b}$ is an
invariant tensor under $H'$, which may be used to raise and lower
indices. The structure constants $f^{\a\b\g}$ are then totally
anti-symmetric.

Taking the covariant derivative of (4.15) it follows
that the covariant derivative of $h^\a$ commutes with
$f^{IJ}$, and must therefore be proportional to the same tensors,
i.e.,
$$
D_i(\Gnul)\,h^\a_{jk}(\f) = \O_{i}^{\a\b} (\f)\, h^\b_{jk}(\f) \,.
\eqno(4.18)
$$
In other words, the tensors $h^\a_{ij}$ are covariantly constant
with respect to the Christoffel connection and some
connection $\O^{\a\b}_{i}$. In view of (4.17) this connection is
anti-symmetric in $\a$ and $\b$.

The fact that $\hat R$ commutes with $SO(N)$ (cf. (4.14)) thus
implies that locally the Riemann tensor can be written as
$$
R_{ijkl} = \ft18\Big\{ f^{IJ}_{ij}\,f^{IJ}_{kl} + C_{\a\b} \,
h_{ij}^\a\,h_{kl}^\b\Big\}  \,, \eqno(4.19)
$$
where $C_{\a\b}(\f)$ is some unknown tensor, symmetric in $\a$
and $\b$, so that the curvature satisfies the pair-exchange
property. According to (4.18-19) and the second equation of
(4.3), the curvature and its multiple
covariant derivatives take their values in the algebra
corresponding $SO(N)\otimes H'$. Therefore the target-space
holonomy group must be contained in this group. Note, however,
that the holonomy group could in principle be smaller than
$SO(N)\otimes H'$, depending on the actual values taken by the
tensor $C_{\a\b}$ and the connection $\O_i^{\a\b}$. It is known
[18] that spaces with restricted holonomy groups have special
properties, so we expect (4.19) to have important consequences.
We shall return to this aspect in section~5.

The fact that we are dealing with an Einstein space implies
$$
C_{\a\b} \, h_{\,i}^{\a\,k}\, h_{kj}^\b = \big[N(N-1)- 8c\big] \,
g_{ij} \,.  \eqno(4.20)
$$
Obviously, the above expression is invariant under $H'$, so that
$$
C_{\d(\a}\,f^{\d\g}_{\b)} \,h^\a h^\b =0\,. \eqno(4.21)
$$

To ensure that the Riemann curvature satisfies the cyclicity
property, the tensors $f^{IJ}$ and $h^\a$ should satisfy
$$
f^{IJ}_{[ij}\,f^{IJ}_{kl]} + C_{\a\b} \, h_{[ij}^\a\,
h_{kl]}^\b =0 \,.   \eqno(4.22)
$$
It is not easy to solve this equation in full generality.
Therefore we first consider its contraction with $f^{KL}_{kl}$
and $h^\a_{kl}$, using (4.5--6) and
$$
f_{IJ}^{ij}\,h^\a_{ij}=0\,. \eqno(4.23)
$$
The latter relation follows from the cyclicity of the trace and
the fact that  (for $N>2$) every tensor $f^{IJ}$ can be written as
the commutator of two such tensors (cf. (4.5)). Note that this is
also in accord with (4.11) and (4.19). For the generators $h^\a$
we used the same argument when imposing (4.17) to ensure
that the trace of the product of two generators belonging to
different factor groups $H''$ vanishes.

The contraction of (4.22) with $f$ leads again to (4.20), while with
$h^\a$ we find
$$
2d_N\,C_\a{}^\b + C_{\g\d}\,f^{\b\g}_\l\,f^{\d\l}_\a -16 c\,
\d^\b_\a=0 \, . \eqno(4.24)
$$
This result shows that $C_{\a\b}$ vanishes when $\a$ and $\b$
belong to the different factor groups of $H'$. For that reason we
may consider (4.24) and (4.21) for the simple subgroups
separately. For the Abelian factor (4.25) can be solved directly,
$$
C_{\a\b}(H'') = {8\,c\over d_N}\d_{\a\b} \qquad \hbox{$\a,\b \in h''$
Abelian} \eqno(4.25)
$$
For the simple factor groups, it is more difficult to find the
solution of $C_{\a\b}$, but after multiplying with $h^\a\,h^\b$,
with $\a$ and $\b$ belonging to the generators of the simple
factor group, and making use of (4.21), we find
$$
C_{\a\b}(H'')\,h^\a h^\b = {16\, c\over 2d_N + c_2(H'')}\; h^\a
h^\a\, ,
\qquad \hbox{with $\a,\b \in h''$ } \eqno(4.26)
$$
where
$$
f^\a{}_{\!\g\d}\, f^{\g\d}{}_{\!\b} = c_2(H'')\, \d^\a_\b \,.
\eqno(4.27)
$$
In the last equation we used Schur's lemma. Observe that (4.26)
applies also to the Abelian factor, as $c_2(H'')=0$ in that case.

Now there is one more conclusion we can draw from (4.22), namely
that the group $SO(N)\otimes H'$ must act {\it irreducibly} on
the target space. To show this, it is convenient to rewrite
(4.22) with target-space indices. Let us then assume that there is
a subspace which is left invariant by $SO(N)\otimes H'$, so that
this group acts reducibly. Denote the indices of this invariant
subspace by $i_\parallel,j_\parallel, \ldots$, and the indices of
its orthogonal complement by
$i_\perp,j_\perp,\ldots$. Subsequently consider the cyclicity
equation (4.22), with indices $i_\parallel,j_\parallel,k_\perp$
and $l_\perp$. Because of the invariance of the subspace there are no
generators with mixed indices, so that (4.22) reduces to
$$
f^{IJ}_{i_\parallel j_\parallel}\,f^{IJ}_{k_\perp l_\perp} +
C_{\a\b} \, h_{i_\parallel j_\parallel }^\a\,
h_{k_\perp l_\perp}^\b =0 \,.   \eqno(4.28)
$$
However, contracting this with $f^{KL}_{k_\perp l_\perp}$ leads
to an immediate contradiction. Hence we conclude that
$SO(N)\otimes H'$ acts irreducibly on the target space.

By Schur's lemma, this shows that the abelian factor in $H'$ has
dimension 0 or 1, with the square of its corresponding
generator $h$ equal to $h^2 = -(2/k)\,\bf 1$. Furthermore both
$C_{\a\b} \, h^\a\, h^\b$ and $h^\a\,h^\a$, with the generators
restricted to one of the factor groups $H''$, are proportional to
the unit matrix. In this way we find
$$\eqalignno{
\big(h^\a\,h^\a\big)_{ij} &= - {2\,\hbox{dim $H''$}\over k}\;
g_{ij} \,,\cr
C_{\a\b}(H'')\,\big(h^\a\,h^\b\big)_{ij} &= - {32\, c\, \over 2d_N +
c_2(H'')}\,{\hbox{dim $H''$}\over k}\; g_{ij} \,,&(4.29)\cr}
$$
where the sum extends over the generators of each of the factor
groups $H''$ separately. Last but not least, as $SO(N)\otimes H'$
leaves the subspace invariant constituted by equivalent $SO(N)$
representations, it follows that the target space should
decompose entirely into $SO(N)$ representations that
are  {\it equivalent}. Consequently, we may put $k_1=k$ and
$k_2=0$.

Now we substitute (4.29) into (4.20) to obtain a relation between
$N$ and the number of supermultiplets. Using that $c_2$ equals
$2(k-2)$, $4k$ and $8(k+1)$, for $SO(k)$, $SU(k)$ and $Sp(k)$,
while the dimensions of these groups are equal to $\ft12 k(k-1)$,
$k^2-1$ and $k(2k+1)$, respectively, leads to the following
equations,
$$
{N(N-1) \over 8\,c}= \cases{
  {\displaystyle{d_N-1\over d_N+k-2}} &for $N=7,8,9$~mod~8\ , \cr
\noalign{\vskip1mm}
  {\displaystyle{d_N^2-4\over d_N(d_N+2k)}} &for $N=6$~mod~4\ ,
\cr
\noalign{\vskip1mm}
  {\displaystyle{d_N+2\over d_N+4k+4}} &for $N=3,5,12$~mod~8\ ,
\cr} \eqno(4.30)
$$
where $c$ was defined in (4.13). From these equations one may
verify that $N(N-1) -d_N$ must be positive, which implies
that there can be no solutions for $N>17$. Therefore it remains
to search for a finite number of explicit solutions, which are
rather rare in view  of the fact that the parameters $N$ and $k$
must be integers. The result of this search is shown in Table~2.

\topinsert
\baselineskip= 4 true mm

$$\vbox{\offinterlineskip
\hrule
\halign{&\vrule# & \strut\quad\hfil#\hfil\quad \cr
height2pt&\omit  &&\omit&&\omit&&\omit&&\omit      & \cr
& $ N $ && $d_N$ && $k$  && $c$       && $H'$      & \cr
height2pt&\omit  &&\omit&&\omit&&\omit&&\omit      & \cr
 \noalign{\hrule}
height2pt&\omit&&\omit&&\omit&&\omit&&\omit      & \cr
& 16  && 128 && 1       && 30            && $\bf 1$     & \cr
& 12  && 64  && 1       && 18            && $Sp(1)$     & \cr
& 10  && 32  && 1       && 12            && $U(1)$      & \cr
&  9  && 16  && 1       && 9             && $\bf 1$     & \cr
&  8  && 8   && $k$     && $6+k$         && $SO(k)$     & \cr
&  6  && 8   && $k$     && $4+k$         && $U(k)$      & \cr
&  5  && 8   && $k$     && $3+k$         && $Sp(k)$     & \cr
&  4  && 4   && $k_\pm$ && $2+k_\pm$     && $Sp(k_\pm)$ & \cr
&  3  && 4   && $k$     && $\ft12(2+k)$  && $Sp(k)$     & \cr
height2pt&\omit &&\omit&&\omit&&\omit&&\omit &\cr}
\hrule}
$$

\narrower\narrower\noindent
Table~2. All solutions to (4.30) with $N=3$ or $N\geq 5$, which
correspond to possible non-linear sigma models coupled to extended
supergravity in terms of $N$ and the number of supermultiplets
$k$. The case $N=4$ is given for comparison. There one can have
two independent quaternionic subspaces corresponding to $k_+$ and
$k_-$ inequivalent supermultiplets.

\endinsert

We should stress that so far we did not determine the tensor
$C_{\a\b}$. An obvious solution is to choose it equal to
$\d_{\a\b}$ for every factor group $H''$. In that case the
Riemann tensor takes its values in the algebra corresponding to
$SO(N)\otimes H'$ (in the spinor representation of $SO(N)$ and
the defining representation of $H'$), and it also invariant under
this group. However, it is possible that there are alternative
solutions for $C_{\a\b}$, corresponding to non-trivial
solutions of (4.24). The Riemann tensor could then take its
values in the algebra
corresponding to a subgroup of $SO(N)\otimes H'$ (which should
still act irreducibly on the target space). Let us denote this
group by $\hat H'$ and assume that it can be written as a
product of subgroups $\hat H''$ that are Abelian (because of
Schur's lemma, the Abelian group is at most one-dimensional) or
simple.  In addition to (4.24) also the following condition must
then be satisfied
$$
\sum_{\hat H''\subset \hat H'} {\hbox{dim $\hat H''$}\over 2d_N
+c_2(\hat H'')}
= \sum_{H''\subset H'} {\hbox{dim $H''$}\over 2d_N
+c_2(H'')} \; ,
\eqno(4.31)
$$
where the subgroups $H''$ are known from Table~2. For an
explicit example of this phenomenon consider $d_N=4$ with the
indices $\a,\b$ taking values in the Lie algebra corresponding to
$Sp(k)$. In that case one
obvious solution corresponds to $C_{\a\b}\propto \d_{\a\b}$,
while a second solutions is obtained by restricting
$C_{\a\b}$ to take only non-zero values for $\a,\b$
corresponding to the generators of the obvious $U(k)$
subgroup. We leave it to the reader to verify that in both cases
one can satisfy (4.24) and (4.31). This example is relevant
for $N=3$, where indeed there exist homogeneous spaces
corresponding to these solutions, namely $Sp(1,k)/(Sp(1)\otimes
Sp(k))$ and $U(2,k)/(U(2)\otimes U(k))$. As we shall discuss in
section~5, the fact that the holonomy group is reduced has
important consequences for the target space.

At this point we have not yet attempted to solve (4.22). The
easiest way to find solutions to this equation is to assume that
one is dealing with a homogeneous space, in which case (4.22) is
just one of the Jacobi identities for the generators of the
isometry group. This will also be discussed in
section~5. For a coset space $G/H$ one expects the Riemann
tensor to take its values in the Lie algebra of $H$. In the case
at hand we know that $H$ must be contained in $SO(N)\otimes H'$.
For a given group $H$ one knows the dimension of $G$, and
in this way it is relatively easy to find coset spaces that
satisfy all the restrictions given above.

Now we turn to a discussion of the $N=4$ theories. An important
role is played by the symmetric tensor $J$, whose definition and
main properties
were given in (4.6--10). As its eigenvalues are equal to $\pm 1$,
we can use it to define the projection operators
$$
\Pi_\pm^{\,i}{}_j = \ft12(\d^i_j\pm J^i_{\,j})\,. \eqno(4.32)
$$
By means of these projectors one decompose the target space into
two subspaces. Because of the fact that the tensors $\Pi_\pm$ are
covariantly constant, the Riemann tensor is only non-vanishing
when all its indices take values in the same subspace (to see
this use the cyclicity of the curvature). Hence we decompose the
curvature into two tensors $R^{(\pm)}_{ijkl}$, satisfying
$$
\Pi_\pm^{\,j}{}_i\, R^{(\mp)}_{jklm} =
\Pi_\pm^{\,j}{}_i\,D_j(\Gnul)\, R^{(\mp)}_{klmn} = 0\, ,\eqno(4.33)
$$
where the second equation follows from the first one combined
with the Bianchi identity. Under these circumstances, the space
is locally a product of two separate Riemannian spaces; this means
that one can choose coordinates such that the metric acquires a
block-diagonal form, in accordance  with the projectors (4.32),
where the metric of one subspace does not depend on the
coordinates of the other one.

Furthermore, because the almost-complex structures commute with
the tensor $J$, they can be decomposed into almost-complex
structures belonging to the two subspaces. Hence we may introduce
two tensors $f_P^{(\pm)\,i}{}_j$, which are only non-zero when
both indices take values in the corresponding subspace, although
at this stage they may still depend on the coordinates of both
subspaces. Decomposing the $SO(4)$ connections in terms of two sets of
$SO(3)$ connections,
$$
Q^{(\pm)\,P}_i =- \ft12 \e^{PQR}\,Q_i^{QR} \mp Q_i^P\, ,
\eqno(4.34)
$$
one can write (3.28) as follows
$$
D_k(\Gnul) \,f_{ij}^{(\pm)\,P} + \e^{PQR} Q^{(\pm)\,Q}_k\,
f_{ij}^{(\pm)\,R} =0\, ,    \eqno(4.35)
$$
while, according to (3.26), the curvatures of the two connections are
equal to
$$
R_{ij}^P(Q^{(\pm)}) =\pa_{i}Q^{(\pm)\,P}_{j} -\pa_{j}Q^{(\pm)\,
P}_{i} +\e^{PQR}\,
Q^{(\pm)\,Q}_i\,Q^{(\pm)\,R}_j =  \pm  f^{(\pm)\,P}_{ij}\,.
\eqno(4.36)
$$
Hence the curvatures $R^P(Q^{(\pm)})$ vanish in the subspace
projected out by $\Pi_\mp$. Therefore by a suitable $SO(3)$ gauge
transformations, one can ensure that the connections $Q^{(\pm)\,P}$
vanish in this subspace. The remaining identities then ensure
that the two spaces decouple completely, with separate complex
structures $f^{(\pm)}$ and connections $Q^{(\pm)}$ with
components in the corresponding subspace and depending only on
the coordinates of the coordinates of that subspace. Note that
the tensors $f^{(\pm)\,P}$ define almost-complex structures in their
respective subspaces. We should perhaps point out here that these
two subspaces do not decouple in the field theory, but interact
via the coupling to the dreibein and gravitino fields.

Hence we may now concentrate on one of these subspaces separately.
Dropping all superscripts $(\pm)$, the geometry in the subspace
is subject to the following equations
$$\eqalignno{
& f_P\,f_Q = -\d_{PQ} \,{\bf 1} \mp \e_{PQR}\, f_R \,,\cr
& D_k(\Gnul) \,f_{ij}^{P} + \e^{PQR} Q^Q_k\, f_{ij}^{R} =0\, ,
 &(4.37)\cr
& R_{ij}^P(Q) =  \pm f^{P}_{ij}\,. \cr}
$$
The subspace transforms under the action of the corresponding
SO(3) group according to inequivalent representations. Again, as
we have three almost-complex structures that are covariantly with
respect to a non-trivial $Sp(1)$ connection, the space is
quaternionic.

For reasons of comparison we repeat the some of the same steps as
in the more general case. Contracting the
integrability condition corresponding to the second equation of
(4.37) with the almost-complex structures and the metric yields
the analogue of (4.11) and (4.12), but with different
normalizations,
$$
R_{ijkl}\,f_P^{kl} = \ft12d_\pm \, f_{P\,ij}\, ,\qquad
R_{ij} = \ft14(8+   d_\pm)\, g_{ij}\,. \eqno(4.38)
$$
where $d_\pm= 4 k_\pm$ is the dimension of the subspace and
$k_\pm$ the number of supermultiplets (which equals the
quaternionic dimension of the subspace). Furthermore we
have a similar decomposition of the curvature as in (4.19),
$$
R_{ijkl} = \ft12 \Big\{ f^{P}_{ij}\,f^{P}_{kl} + C_{\a\b} \,
h_{ij}^\a\,h_{kl}^\b\Big\}  \,, \eqno(4.39)
$$
where the tensors $h^\a$, together with the identity,  span the
centralizer of the almost-complex
structures, so that they generate the group $Sp(k_\pm)$.
Together with the complex structures they generate the group
$Sp(1)\otimes Sp(k_\pm)$, which must again act irreducibly. Again
one derives
$$
C_{\a\b}\,(h^\a h^\b)_{ij} = -\ft12(2+d_\pm)\, g_{ij}\,.
\eqno(4.40)
$$

We should point out that the presence of the two separate
quaternionic spaces can be understood from $N=2$ supergravity in
four space-time dimensions. In that case there exist two
inequivalent matter multiplets. The vector multiplets, whose
scalar fields parametrize a K\"ahler manifold [27], and the
scalar (or hyper-)multiplets, whose scalar fields parametrize a
quaternionic manifold [28]. Upon dimensional reduction the
K\"ahler space of the vector multiplets is converted into a
quaternionic space (although not the most general) [29],
so that one obtains two quaternionic
spaces associated with inequivalent supermultiplets.

Perhaps we should explain why this phenomenon can only happen
for $N=4$, while there are inequivalent multiplets for all values
$N=4$~mod~4, as we showed in section~2. The reason is that the
group $SO(N)\otimes H'$ must act irreducibly in the target space,
so that only one type of multiplet is allowed. The sitution for
$N=4$ is different, because the group $SO(4)$ factors into two
separate $SO(3)$ groups, each of them acting in a different
subspace of the target space.

The question that remains to be answered is what the possible
spaces are corresponding to $N>4$. As we shall argue in the next
section, it turns out that these spaces are unique. After
identifying each one of them it is rather straightforward to
verify that all equations of this section are indeed satisfied.

\beginsection 5. Homogeneous spaces

A striking feature of the
results derived in the foregoing section is that, except for the
low values $N\leq 4$, the number of possible theories is rather
limited. In particular, for $N>8$, there remain only four theories
based on a single supermultiplet
corresponding to $N=9,10,12$ and 16. The bound $N\leq 16$ was obtained
here solely on the basis of mathematical considerations; since there is
no helicity in three dimensions, we cannot rely on ``physical"
arguments, unlike in four space-time
dimensions, where the analogous bound $N\leq 8$ follows
from requiring absence of massless states of helicity higher than 2.
The arguments of section 4 are not yet strong enough to determine
the target manifolds, since we used only a contracted version of (4.22);
to find out what the possible spaces are, one must exploit the full
content of these identities. Fortunately, we can now invoke a powerful
mathematical theorem to prove that the target spaces
are, in fact, symmetric and therefore homogeneous for sufficiently
high $N$.

\medskip

\noindent
{\bf Theorem} [19]: {\it Let $\cal M$ be an irreducible Riemannian
manifold. If the holonomy group at a point $p \in {\cal M}$
does not act transitively on the unit sphere in the tangent space
$T_p {\cal M}$ at $p$, then $\cal M$ is a symmetric space of
rank $\geq 2$.}
\medskip

\noindent
The content of this theorem can be rephrased as follows: if the
holonomy group of $\cal M$ is sufficiently ``small" with respect to
the generic holonomy group (i.e. $SO(d)$ for an arbitrary
$d$-dimensional Riemannian manifold), then the manifold is completely
determined; if, on the other hand, it is
``large", then little can be said, and there is a greater variety of
spaces. We note, however, that the possible holonomy groups for
irreducible non-symmetric Riemannian manifolds cannot be arbitrary
subgroups of $SO(d)$, but are strongly restricted; a complete list
is given in Corollary 10.92 of [18].
In the case at hand, all the necessary information is
encoded in the explicit formula (4.19) for the curvature
tensor, which tells us that the holonomy group is contained in
$SO(N)\otimes H'$, where the centralizer subgroup $H'$ can be read
off from Table 2. As the dimension of target space is $d= k d_N$,
we must therefore check whether or not the group
$SO(N)\otimes H'$ acts transitively on the unit sphere $S^{d-1}$.
When it does not, then the holonomy group $SO(N)\otimes \hat H$,
which is contained in it,
does not act transitively either and we can apply
the theorem. This allows us to understand the
limitations on the number of possible theories from a slightly
different point of view: extended supergravity theories are scarce
because the mismatch between the actual holonomy group
$SO(N)\otimes \hat H$ and the generic holonomy group $SO(d)= SO(k d_N)$
becomes too big for $N>4$. For $N\leq 4$, the information provided
by (4.19) is not sufficient to completely determine
the manifold. In particular, for $N=1$, there are no restrictions
at all, and the target space is an arbitrary Riemannian
manifold. For $N=2$, the holonomy group has a $U(1)$ factor; since
there is one complex structure, the manifold must be K\"ahler, and
the holonomy group is contained in
$U(k)$ with $d=2k$. As this group acts
transitively on the sphere $S^{2k-1}$, we get no further restrictions
from the theorem. For $N=3$ and $N=4$, the target spaces are
quaternionic manifolds of dimension $d=4k$ and $d_\pm = 4k_\pm$,
respectively, and the holonomy
group is contained in $Sp(1)\otimes Sp(k)$. Since the group
$Sp(1)\otimes Sp(k)$ acts transitively on the sphere $S^{4k-1}$,
the theorem imposes no immediate restrictions on
the manifold. For all higher values of $N$ with the exception of
$N=9$, the group $SO(N)\otimes H'$, and therefore the
holonomy group does {\it not} act transitively. According to the theorem
we can then uniquely determine the possible target manifolds
by matching the values of $N$ and $d$ with the list of symmetric
spaces. This identification leads to the list of spaces shown in
Table~3, which forms a central result of this
paper.\footnote{${}^5$}{By some abuse of notation we
wrote orthogonal groups for the cosets where possible.
It should be clear from the text in section~4
what the representations are in which the isotropy group
acts. As $SO(N)$ acts in the spinor representation it would be
appropriate to denote is as $Spin(N)$,
whereas the $SO(3)$ group for $N=12$ is actually
$Sp(1)$. Observe the
importance of triality for the $N=8$ coset space,
which can be used to interchange
vector and spinor representations of $SO(8)$.}
All non-linear
sigma models coupled to $N\geq5$ supergravity
are thus uniquely determined.
The maximal number of supersymmetries is
$N=16$, which corresponds to the theory
constructed quite some time ago in [12].
The case $N=9$ may seem special, as $Spin(9)$ does act
transitively on $S^{15}$, but it can be shown that the coset space
$F_4/Spin(9)$ (which is of rank 1) is the only
solution [30].\footnote{${}^6$}{In [18], the reader may
find the list
of subgroups of $SO(d)$ which act transitively on $S^{d-1}$. Besides
the regular groups, there are three exceptional cases, namely $G_2$
acting on $S^6$, $Spin(7)$ on $S^7$ and $Spin(9)$ on $S^{15}$. The
first two of these play no role in our analysis, because the
associated manifolds are Ricci flat [18], which would lead to a
contradiction with (4.12) and (4.13).}

\topinsert
\baselineskip = 4 true mm
$$\vbox{\offinterlineskip
\hrule
\halign{&\vrule# & \strut\quad\hfil#\hfil\quad \cr
height2pt&\omit  &&\omit&&\omit&&\omit&&\omit&&\omit     & \cr
& $ N $ && $d_N$ && $k$  && $c$       &&$G/H$ && rank   & \cr
height2pt&\omit  &&\omit&&\omit&&\omit&&\omit&&\omit     & \cr
 \noalign{\hrule}
height2pt&\omit  &&\omit&&\omit&&\omit&&\omit&&\omit     & \cr
& 16  && 128  && 1   && 30            &&$E_{8(+8)}/SO(16)$
                   && 8 & \cr
& 12  && 64   && 1   && 18
              &&$E_{7(-14)}/(SO(12)\otimes SO(3))$ && 4 & \cr
& 10  && 32   && 1   && 12
              &&$E_{6(-14)}/(SO(10)\otimes SO(2))$ && 2 & \cr
&  9  && 16   && 1   && 9             &&$F_{4(-20)}/SO(9)$
                   && 1 & \cr
&  8  && 8    && $k$ && $6+k$
              &&$SO(8,k)/(SO(8)\otimes SO(k))$ && max$\,(8,k)$ & \cr
&  6  && 8    && $k$ && $4+k$
              &&$SU(4,k)/S(U(4)\otimes U(k))$  && max$\,(4,k)$ & \cr
&  5  && 8    && $k$ && $3+k$
              &&$Sp(2,k)/(Sp(2)\otimes Sp(k))$ && max$\,(2,k)$ & \cr
height1pt&\omit &&\omit&&\omit&&\omit&&\omit&&\omit &\cr}
\hrule} $$
\narrower\narrower\noindent
Table~3. Complete list of target spaces for $N\geq 5$ supergravity
theories. The coefficient $c$, defined in (4.12), coincides with
the dual Coxeter number of the group $G$.
\endinsert

We expect that the theories with even $N$ in Table~3
can be obtained by dimensional reduction of the corresponding
$N/2$ theories in four space-time
dimensions. To obtain the theories
with odd $N$, one would have to further truncate the dimensionally
reduced theories, but, evidently, neither the target spaces nor the
fact that there are no theories for certain odd values of $N$ below
$N=16$ and none at all above $N=16$ could have been reliably
predicted on the basis of such arguments. We should perhaps point out
that exceptional groups (including $G_2$) also appear for symmetric
quaternionic spaces. All homogeneous
quaternionic spaces are known and were
given in [31] (see also [23]).

Having established that the target spaces are symmetric for
sufficiently high $N$, we devote the remainder of this section
to elucidating some features of the relation between the
target-space formulation of locally supersymmetric theories as
given in section 3 and the formulation of extended supergravity
theories as $G/H$ coset space theories (see, for instance,
[32,12]). In particular we shall indicate how
some of the results of our work arise in the context of the
latter formulation. We assume, in accord with the spaces listed in
Table~3, that $G$ is a non-compact group and $H$ its maximal
compact subgroup, so that the space is symmetric. For $N\geq5$
the possible choices for $G$ and $H$ can be gleaned from Table~3,
but our results can be applied for other cases as well.
Together with the results derived in section~3, this information
then gives an explicit representation of the
Lagrangian and supersymmetry transformations of the theory.

Let us first discuss the group-theoretical aspects in a little
more detail. From section~4 we know that
the group $H$ always factorizes according to
$SO(N)\otimes \hat H$, where $\hat H\subset H'$ (for the spaces
listed in Table~3, $\hat H$ and $H'$ do actually coincide).
The generators of the group $\hat H$ will be denoted by $h^\a$
where the indices $\a$ now take their values in the Lie algebra
of $\hat H$: $\a =1,\ldots, \hbox{dim} \hat H$.  They commute with
fermion number and with the matrices $\G^I_{A\dot A}$,
$$
h^\a_{AC} \G^I_{C \dot B} + h^\a_{\dot B \dot C} \G^I_{A \dot C}
= 0 \,.                \eqno (5.1)
$$
Denoting the $SO(N)$ generators by $X^{IJ} = - X^{JI}$, where
$I,J,\ldots=1,\ldots,N$, and the remaining (coset) generators by
$Y^A$, where the boson indices $A,B,\ldots$ (or the fermionic
ones $\dot A,\dot B ,\ldots$) $=1,\ldots,d$ were already
introduced in section~2, the Lie algebra of $G$ is characterized
by the commutation relations
$$
\eqalign{
&\big[ X^{IJ} , X^{KL} \big] = \d^{JK} X^{IL} - \d^{IK} X^{JL} -
         \d^{JL} X^{IK} + \d^{IL} X^{JK} \,,    \cr
&\big[ X^\a , X^\b \big] = f^{\a \b}{}_{\!\g}\, X^\g \; \; ,
\quad \qquad
 \big[ X^{IJ} , X^\a \big] = 0  \,, \cr
&\big[ X^{IJ} , Y^A \big] = - \ft12 \G^{IJ}_{AB} \,Y^B  \; ,
\qquad \big[ X^\a , Y^A \big] = - h^\a_{AB} \,Y^B \,, \cr
&\big[ Y^A , Y^B \big] = \ft14 \G^{IJ}_{AB} \,X^{IJ} +
   \ft18 C_{\a \b} \, h^\a_{AB} \,X^\b \,,    \cr}     \eqno (5.2)
$$
where $\G^{IJ}_{AB} \equiv \G^{[I}_{A \dot A}\G^{J]}_{B \dot A}$,
so that $\ft12\G^{IJ}_{AB}$ generates the spinor representation
of $SO(N)$. Likewise
$$
h^\a_{AC}\, h^\b_{CB}- h^\b_{AC}\, h^\a_{CB} =f^{\a\b}{}_{\!\g}
 \, h^\g_{AB} \,.\eqno(5.3)
$$
The tensor $C_{\a \b}$ coincides with the tensor introduced in
(4.19). Most of the Jacobi identities implied by the algebra (5.2)
are trivially satisfied once we assume that $C_{\a\b}$ is $\hat
H$ invariant. The remaining identity, and the one
that leads to the most stringent constraints on $G$, arises from
the commutator $[[Y^A , Y^B ], Y^C ]$; it reads
$$
\G^{IJ}_{[AB}\, \G^{IJ}_{CD]} + C_{\a \b}\, h^\a_{[AB} \,
h^\b_{CD]} = 0\,. \eqno (5.4)
$$
This equation is just (4.19), except that $C_{\a\b}$
is now assumed to be $\hat H$
invariant. From section~3 we can therefore deduce its values for
the spaces listed in Table~3, using the normalization
(4.17). For $N=16$ and 9, $C_{\a\b}$ obviously
vanishes; for $N=12$, 10, 8 and 5, $\hat H$ is simple, so that
$C_{\a\b}$ is proportional to the identity, and its eigenvalues
are equal to 2, 3, 8 and 2, respectively. The case $N=6$ is slightly
more complicated. For the $SU(k)$ subgroup $C_{\a\b}$ is proportional
to the identity with eigenvalue equal to 4, whereas for the
$U(1)$ subgroup, we have the eigenvalue $4+k$.
In the appendix, we will give an explicit proof of the Jacobi identity
(5.4) for the groups $E_8 , E_7 , E_6$ and $F_4$.

In the coset space formulation
the scalar fields that parametrize the coset space
are characterized  by a matrix
$\cV (x) \in G/H$, on which $G$ acts as a rigid symmetry
group from the left, while $H$ is realized as a local symmetry
acting from the right. To understand that this description is
equivalent to the one in terms of the target-space coordinate
fields $\phi^i (x)$, we note that the matrix $\cV$ represents
$d = {\rm dim}(G/H) = {\rm dim} G - {\rm dim} H$ physical degrees
of freedom. The spurious (gauge) degrees of freedom associated with
the subgroup $H$ can be eliminated by choosing a special (``unitary")
gauge where the matrix $\cV$ is directly parametrized through the
target-space coordinates $\phi^i (x)$ used before, i.e.
$\cV = \cV (\phi^i (x))$. To maintain
this gauge choice under local supersymmetry transformations
compensating $H$ rotations will be needed. We will
also need a vielbein $e_i^A$ as well as gauge connections
$Q_i^{IJ}$ and $Q_i^\a$ for the tangent-space group
$SO(N)\otimes \hat H$.
These are defined by (for a systematic and rather complete
discussion of coset spaces, see e.g. [33])
$$
{\cV}^{-1} \partial_i \cV = \ft12  Q_i^{IJ}\, X^{IJ}
          + Q_i^\a\, X^\a   + e_i^A \,Y^A  \,,\eqno (5.5)
$$
where $\pa_i$ is the derivative with respect to the target-space
coordinate $\phi^i$.

The integrability condition corresponding to (5.5) are the
so-called Cartan-Maurer equations. In this case they read
$$\eqalignno{
D_{[i} e_{j]}^A &= \pa_{[i} e_{j]}^A  +
  \big( \ft14 Q_{[i}^{IJ} \,\G^{IJ}_{AB} +
    Q_{[i}^\a \,h^\a_{AB} \big) e_{j]}^B = 0 \,,&(5.6)\cr
R_{ij}^{IJ} &= -\ft12 e^A_i\,e^B_j \, \G_{AB}^{IJ} \,,
 &(5.7)\cr
R_{ij}^\a &=- \ft18 e^A_i\,e^B_j\, C_{\a\b} \, h^\b_{AB}  \,,
&(5.8) \cr}
$$
where $R_{ij}^{IJ}$ was already defined in (4.3), while
$R_{ij}^\a$ equals
$$
R_{ij}^\a \equiv \pa_i Q_j^\a -
 \pa_j Q_i^\a +f^\a{}_{\!\b \g }\, Q_i^\b\, Q_j^\g  \,.  \eqno(5.9)
$$

The  geometrical content of the theory is fixed once we identify
$e^A_i$ as the vielbein
of the coset manifold with $Q_i^{IJ}$ and $Q_i^\a$ the spin-connection
fields. The latter take their values in the algebra of the
isotropy group, which is the
subgroup of $SO(d)$ that acts on the tangent space with
the generators $\ft12\G^{IJ}$ and $h^\a$ defined above.
According to (5.5) the space is torsion-free, so that the
vielbein is covariantly constant with respect to the
Christoffel connection,
$$
D_i e_j^A = \pa_i e_j^A - \Gnul^k_{ij} e_k^A +
  \big( \ft14 Q_i^{IJ} \G^{IJ}_{AB} +
   Q_i^\a h^\a_{AB} \big) e_i^B = 0  \eqno (5.10)
$$
The vielbein $e_i^A$ is related to the
target-space metric of the preceding section by
$$
 g_{ij} (\phi ) = e_i^A (\phi)\,e_j^B(\phi )\,\eta_{AB}\,,\eqno (5.11)
$$
where $\eta_{AB}$ is a symmetric and $\hat H$-invariant tensor; in case
there is more than one invariant tensor, the metric is thus no longer
unique. The vielbein can also be used to convert curved into flat
indices in the usual fashion;
for instance, the generators of $\hat H$ are related to (a subset
of) the matrices $h^\a_{ij}$ used previously (see (4.15)) by
$$
h^\a_{ij} = h^\a_{AB} e_i^A e_j^B  \,.          \eqno (5.12)
$$
The curvature tensor on $G/H$ can be computed from
$$
R_{ijkl} =- e_k^A\, e_l^B \left(\ft14 R_{ij}^{IJ} \G^{IJ}_{AB} +
        R_{ij}^\a \,h^\a_{AB} \right) \,.   \eqno (5.13)
$$
Using (5.7--8) one thus obtains
$$
R_{ABCD} = {\textstyle {1\over 8}} \Big( \G^{IJ}_{AB}\, \G^{IJ}_{CD} +
   C_{\a \b} \,  h^\a_{AB} \,h^\b_{CD} \Big)\,, \eqno (5.14)
$$
which precisely coincides with (4.39).
In terms of flat indices, the curvature tensor is therefore constant;
moreover, the Jacobi identity (5.4) ensures the cyclicity of the
Riemann tensor and is thus equivalent to (4.22).

{}From the previous sections we know of the existence of
$N-1$ almost-complex structures
$f^P_{ij}$ (remember that $P,Q,\ldots=2,\ldots,N$). In the coset
formulation they can be represented by
$$
f^P_{ij} = \pm \big( \G^P \G^1 \big)_{AB}\, e_i^A\, e_j^B  \,,
  \eqno (5.15)
$$
and are not $SO(N)$ covariant.
On the other hand, the antisymmetric tensors $f^{IJ}_{ij}$, which
were defined in(4.1), are $SO(N)$ covariant, and take the form
$$
f^{IJ}_{ij} = - \G^{IJ}_{AB}\,e_i^A\, e_j^B\,.   \eqno (5.16)
$$
The tensors $f^P_{ij}$ are only almost-complex structures;
from (5.10) and the definition (5.15),
we immediately deduce that
$$
D_i (\Gnul ) f^P_{jk} = \pm \ft14 Q_i^{IJ} \, \big[
\G^{IJ} , \G^P \G^1 \big]
  = - Q_i^Q f^{PQ}_{jk} - Q_i^{PQ} f^Q_{jk}\,,     \eqno (5.17)
$$
where we made use of the definition (4.2). Relation (5.17) is
nothing but the previous formula (3.28).

In the coset formulation the fermion fields do not carry
target-space indices. To appreciate this feature,
let us recall the supersymmetry transformation
$$
\d\f^i = \ft12 \big(\bar\e\,\x^i + \bar\e{}^P f^{\,i}_{Pj}
\x^j\big)\,.\eqno(5.18)
$$
By making use of the supersymmetry transformation with
parameter $\e$, one naturally defines fermion fields
that transform as the components of a target-space vector.
In the coset formulation, on the other hand,
one considers $\cV^{-1}\d\cV$, which takes
its values in the Lie algebra of $G$. By a suitable
(field-dependent) $H$ transformation, this expression
can be restricted to take its values in the generators
$Y^A$. This motivates one to introduce
fermion fields $\x^{\dot A}$
that transform covariantly under $H$, so that the supersymmetry
variation takes the form
$$
\cV^{-1} \delta \cV = \ft12 \bar \e^I \x^{\dot A}\,
\G^I_{A\dot A} \;Y^A \,.   \eqno (5.19)
$$
In a given gauge the two transformations should coincide,
modulo a compensating (field-dependent) $H$ transformation
to maintain the gauge choice

By comparing the two supersymmetry variations we can find the
relation between the fermion fields $\x^i$ and $\x^{\dot A}$.
We first observe that the direct variation of $\cV$ yields
$$
\cV^{-1} \delta \cV = \delta \phi^i \, \cV^{-1} \partial_i \cV
= \delta \phi^i \, \Big( \ft12 Q_i^{IJ} \,X^{IJ} +Q^\a_i\,X^\a
+ e_i^A \,Y^A \Big)\,. \eqno (5.20)
$$
Obviously the first two terms correspond to infinitesimal
field-dependent $H$ transformations. The last term should be
matched with (5.19), so that
$$
\big(\bar\e\,\x^i + \bar\e{}^P\,f^{\,i}_{Pj}
\x^j\big)\,e_i^A
= \bar \e^I \x^{\dot A}\,
\G^I_{A\dot A} \,.   \eqno (5.21)
$$
Making use of (5.15) this relation leads to the
identifications $\e^1=\pm\e$ and
$$
\x^{\dot A} = \pm \G^1_{A\dot A}\, e_i^A\,\chi^i  \,.   \eqno (5.22)
$$
With this result the variations (5.18) and (5.19) coincide,
provided one adds a compensating $H$
transformation to (5.19) with parameters
$$
\omega^{IJ} =  \delta \phi^i\, Q_i^{IJ}  \, ,\qquad
\omega^\a = \delta \phi^i\, Q_i^\a \,.     \eqno (5.23)
$$
This compensating transformation
must be included in all supersymmetry variations.
To see the corresponding relation for the fermions
$\chi^i$ is slightly more subtle. Using (5.22) we find
$$\eqalignno{
\d \x^{\dot A} & =
\pm \G^1_{A\dot A} \big( \d e_i^A\,\chi^i
+ e_i^A \,\d \chi^i \big) \cr
&= \pm \G^1_{A\dot A}\, \d\f^j\,\big(\pa_j e_i^A- \G^k_{ji}\,e_k^A\big)
 \chi^i  + \ft12\G^I_{\dot A A}\,e_i^A\,\hat{\pa\slash}\f^i \e^I \,,
&(5.24)\cr}
$$
where we made use of (3.25) and (5.15). Here it is important that
$\G^k_{ji}$ is {\it not} the Christoffel connection, but the
modified connection defined in (3.27). Using (5.10) and (3.27)
shows that the first term is equal to
$$
\d_\o\x^{\dot A} = -\ft12\o^{IJ}\,\G^{IJ}_{\dot A\dot B}\,\x^{\dot B}
- \o^\a\,h^\a_{\dot A\dot B}\,\x^{\dot B} \,,\eqno(5.25)
$$
where $h^\a_{\dot A\dot B}= \G^1_{\dot A A}\,h^\a_{AB}\,
\G^1_{\dot B B}$ by virtue of (5.1). In deriving this, we also
made use of (4.2) and (5.10). The terms (5.25) are
precisely cancelled by the compensating transformation (5.23).
The remaining variation thus takes the form
$$
\delta \x^{\dot A} = \ft12 \g^\mu \e^I\, \G^I_{A\dot A} \,
P_\mu^A \,, \eqno (5.26)
$$
where we use the notation
$$
P_\mu^A \equiv \partial_\mu \phi^i \, e_i^A\,.       \eqno (5.27)
$$
Finally, by similar manipulations as described above, one may
verify that
$$\eqalignno{
D_\mu(\G) \x^i & =\pm \big(D_\m(\Gnul)e^i_A\big)
\, \G^1_{A\dot A}\,\x^{\dot A}
\pm e^i_A\,  \G^1_{A\dot A} \,\pa_\m\x^{\dot A}
- e^i_A\,\G^P_{A\dot A}\,Q_j^P\,\pa_\m\f^j   \cr
&= \pm e^i_A\,\G^1_{A\dot A}\Big(
\delta_{\dot A \dot B} \partial_\mu
+ \ft14 Q_\mu^{IJ} \,\G^{IJ}_{\dot A \dot B} +Q^\a_\m\,
h^\a_{\dot A\dot B} \Big)
\x^{\dot B} \,,&(5.28)\cr}
$$
where
$$
Q_\mu^{IJ} = \partial_\mu \phi^i Q_i^{IJ}  \; , \qquad
Q_\mu^\a = \partial_\mu \phi^i Q_i^\a  \; .
    \eqno (5.29)
$$
The modification of the fermionic connection as given in (3.27)
is thus indispensable for recasting the results in such a
systematic and covariant form in the coset formulation.
The reader is advised to consult [12] to see that these
various ingredients are indeed present for the theories
constructed in that work.

\beginsection Appendix

In this appendix, we will establish the crucial Jacobi identity (5.4)
for the exceptional groups $E_8$, $E_7$, $E_6$ and $F_4$.
For the convenience of the reader, we here repeat formula (5.4)
for $C_{\a \b} = A \,\d_{\a \b}$
$$
\GIJ_{[AB} \,\GIJ_{CD]} + A \, h^\a_{[AB} \,h^\a_{CD]} = 0 \,.
\eqno (A.1)
$$
For $G=E_8$ and $F_4$, the subgroup $\hat H$
is trivial, and the
second term is therefore absent. For $G=E_7$ and $G=E_6$, we have
$\hat H = Sp(1)$ and $\hat H= U(1)$,
respectively, so the second term in (A.1)
must be taken into account; with the normalization adopted in
(4.17), we find $A=2$ for $E_7$ and $A=3$ for $E_6$, as stated
below (5.4). To prove (A.1), we will need to know
the Fierz identities for matrices acting on the $d$-dimensional
chiral spinor representations of $SO(N)$ (there is only one
multiplet, so we have $d=d_N$). Since we are dealing with
a $real$ representation of the Clifford algebra, the standard Fierz
identities for complex $\G$-matrix algebras must be modified.
Fierz identities for real Clifford algebras have been derived in
[22]; however, these are not quite suitable for our purposes, and
we will therefore present an alternative formulation.
We will make use of the standard definition
$$
\G^{I_1\cdots I_{2k}} \equiv
\G^{[I_1}\cdots \G^{I_{2k}]} \,.       \eqno (A.2)
$$
Notice that we consider only matrices built out of an $even$
number of $\G$-matrices, which do not mix the $d$-dimensional
chiral subspaces. For brevity, we will denote these matrices by
$\G^{(2k)}$ below, so that $\G^{(2k)}_{AB} \equiv
\G^{I_1 \cdots I_{2k}}_{AB}$. The matrices $\G^{(2k)}$ are symmetric
for even $k$, and antisymmetric for odd $k$.
Let us first record the important formulas
$$
{\rm Tr}\, \big( \Gamma^{ I_1 \cdots I_{2k}}\, \Gamma_{J_1 \cdots
J_{2k}} \big) =
d\,(-)^k\, (2k)!\,  \delta^{I_1 \cdots I_{2k}}_{J_1 \cdots
J_{2k}} \,,\eqno (A.3)
$$
and
$$
\Gamma^{IJ} \,\Gamma^{K_1 \cdots K_{2p}} \, \Gamma^{IJ} =
\big( N - (N-4p)^2  \big)\, \Gamma^{K_1 \cdots K_{2p}}\,,
\eqno (A.4)
$$
which are valid for arbitrary $N$ (traces are understood to be
over the chiral subspace labeled by the indices
$A,B,\ldots= 1,\ldots,d$). From the explicit representation
of the $\G$-matrices in section 2, it is not difficult to check that,
for $N=4$ mod 4, the matrix $\tilde \G$ in (2.5) can be taken equal
to the identity matrix. Since the fermion number operator $\bf F$
is also unity in the chiral subspace, the matrices $\G^{(2k)}$ and
$\G^{(N-2k)}$ are related to each other by duality, hence linearly
dependent; for $2k= N/2$ there are thus only $\half
{N \choose 2}$ linearly independent matrices.
For $N=2$ mod 4, we find $\tilde \G = e$; therefore,
duality now relates $\G^{(N-2k)}$ and $ e \G^{(2k)}$. For odd $N$,
on the other hand, all matrices are linearly independent.

For $\underline{N=8n}$, we have $d=2^{4n-1}$ from Table~1.
Elementary counting arguments show that the matrices ${\bf 1}, \HA ,
\HB , \ldots , \G^{(4n)}$ form a complete and linearly independent
set of (real) $d\times d$ matrices (for the matrices $\G^{(4n)}$,
one must not forget to take into account the self-duality
constraint, as we just explained).
The relevant Fierz identity for an antisymmetric matrix $M_{AB}$
(which is all we need for (A.1)) therefore reads
$$
M_{AB} = - {1 \over d} \sum_{k=1,3,\ldots,{2n-1}}
{1 \over {(2k)!}} \,
      \G^{(2k)}_{AB} \, {\rm Tr} \left( M \G^{(2k)} \right)\,.
                     \eqno (A.5)
$$
Summation over the $2k$ indices $I_1, \ldots , I_{2k}$ is
implied in (A.5) and similar formulas below. For $N=16$, this
sum evidently contains only two terms. Evaluating (A.5) for
the matrix $M_{AB} = \G^{IJ}_{C[A} \G^{IJ}_{B]D}$, we obtain
$$
\G^{IJ}_{C[A} \G^{IJ}_{B]D} =
     {\textstyle  {1 \over {128}} {1 \over {2!}} }
   \HA_{AB} \big( \GIJ \HA  \GIJ \big)_{CD} +
 {\textstyle  {1 \over {128}} {1 \over {6!}}}
   \HC_{AB} \big( \GIJ \HC  \GIJ \big)_{CD} \,.\eqno (A.6)
$$
{}From (A.4), we get $\GIJ \HA  \GIJ = - 128 \HA $ and
$\GIJ \HC \GIJ = 0$, so (A.6) reduces to
$$
\GIJ_{C[A} \GIJ_{B]D} = - \half \GIJ_{AB} \GIJ_{CD} \,,\eqno (A.7)
$$
from which the desired relation (A.1) follows directly (with $A=0$).

For $\underline{N=4+8n}$, we have $d= 2^{2+4n}$. In contrast to
the previous case, a complete set of real $d\times d$ matrices
now cannot be constructed from the $\G$-matrices alone, as one can
quickly verify by counting the number of such matrices.
In addition, however, there are now three complex structures
represented by the antisymmetric matrices $h^\a_{AB}$ for $\a = 1,2,3$,
which generate the centralizer subgroup $Sp(1)$. With the
normalization (4.17), we have $(h^\a)^2 = -2$ (no summation over
$\a$) and
$$
\big[ h^\a \, , \, h^\b \big] = 2 \sqrt{2} \e_{\a \b \g} h^\g \,.
                                    \eqno (A.8)
$$
A complete and linearly independent set of antisymmetric matrices is
given by $h^\a$, $\HA$, $h^\a  \HB , \dots , h^\a \G^{(4n)},
\G^{(4n+2)}$, while the symmetric matrices are
$\bf 1$, $h^\a  \HA $, $\HB  , \dots , \G^{(4n)},$ \break
$h^\a \G^{(4n+2)}$. Instead of writing down the general formula,
let us immediately specialize to $N=12$, so that $d=64$; in this
case, the relevant identities are
$$\eqalignno{
\GIJ_{C[A} &  \GIJ_{B]D} = &(A.9)\cr
  =  & {\textstyle {1 \over {64}}}
\Big\{  \half h^\a_{AB} \big( \GIJ h^\a \GIJ \big)_{CD} +
  \half \HA_{AB} \big( \GIJ \HA  \GIJ \big)_{CD} +   \cr
& \qquad+ {\textstyle {1 \over {4!}}} \big( h^\a \HB  \big)_{AB}
\big( \GIJ h^\a \HB  \GIJ \big)_{CD} +
   {\textstyle  {1 \over {2 \cdot {6!}}} } \HC_{AB} \big( \GIJ
    \HC  \GIJ \big)_{CD} \Big\}    \cr
= & {\textstyle {1 \over {64}}} \Big\{ - 66 h^\a_{AB} h^\a_{CD}
    -26 \GIJ_{AB} \GIJ_{CD} -
          {\textstyle {1 \over 6}} \big( h^\a \HB \big)_{AB}
       \big( h^\a \HB  \big)_{CD} +
{\textstyle {1 \over {120}}} \HC_{AB} \HC_{CD} \Big\}\,, \cr }
$$
and
$$\eqalignno{
h^\a_{C[A} &  h^\a_{B]D} = &(A.10)  \cr
  =   &  {\textstyle {1 \over {64}}}
\Big\{ \half h^\b_{AB} \big( h^\a h^\b h^\a \big)_{CD} +
  \half \HA_{AB} \big( h^\a \HA h^\a \big)_{CD} +   \cr
&\qquad + {\textstyle {1 \over {4!}}} \big( h^\b \HB \big)_{AB}
\big( h^\a  h^\b \HB h^\a \big)_{CD} +
  {\textstyle {1 \over {2\cdot 6! }}} \HC_{AB} \big( h^\a
    \HC  h^\a \big)_{CD} \Big\}   \cr
= &   {\textstyle {1 \over {64}}} \Big\{  h^\a_{AB} h^\a_{CD}
    - 3 \GIJ_{AB} \GIJ_{CD} +
         {\textstyle {1 \over {12}}} \big( h^\a \HB
    \big)_{AB} \big( h^\a \HB \big)_{CD} -   {\textstyle
    {1 \over {240}}} \HC_{AB} \HC_{CD} \Big\}\,, \cr }
$$
where (A.3) was used (the extra factor of $\half$ in front of the
terms containing $\HC$ is due to the self-duality constraint, which
was explained above). It is now straightforward to check that
$$
\GIJ_{C[A} \GIJ_{B]D} + 2 h^\a_{C[A} h^\a_{B]D}
 = - \half \left( \GIJ_{AB} \GIJ_{CD} + 2 h^\a_{AB} h^\a_{CD}
 \right) \,,       \eqno (A.11)
$$
so that (A.1) is satisfied with $A=2$.

For $\underline{N=2+8n}$, we read off $d= 2^{1+4n}$ from Table~1.
There is now only one complex structure represented by the
antisymmetric matrix $h_{AB}$, which generates the group
$U(1)$ and is again normalized such that
$(h)^2 = -2$. The antisymmetric matrices are $h, \HA , h \HB ,
\ldots , \G^{(4n-2)} , h \G^{(4n)}$, while the symmetric ones are $
{\bf 1}, h \HA , \ldots , h \G^{(4n-2)}, \G^{(4n)}$. One checks
that altogether there are $\quart d^2$ antisymmetric and $\quart d^2$
symmetric matrices, so it would seem that we cannot generate a
complete set of matrices in this way. However, we now recall
that the representations are $complex$ for these values of $N$
(see the discussion in section~2), which means that, instead of
getting $d^2$ real matrices, we should end up with $({d\over 2})^2$
complex (i.e. $({d\over 2})^2$ hermitean and $({d\over 2})^2$
anti-hermitean) matrices; this is precisely the number of matrices just
obtained. Specializing to $N=10$ with
$d=32$, the relevant identities read
$$\eqalignno{
\GIJ_{C[A} & \GIJ_{B]D}  =   \cr
    = & \, {\textstyle {1 \over {32}}}
\Big\{ \half h_{AB} \big( \GIJ h \GIJ \big)_{CD} +
  \half \HA_{AB} \big( \GIJ \HA  \GIJ \big)_{CD}
 + {\textstyle {1 \over {4!}}} \big( h \HB  \big)_{AB} \big( \GIJ
    h \HB  \GIJ \big)_{CD}  \Big\}    \cr
= & \, {\textstyle {1 \over {32}}} \Big\{  - 45 h_{AB} h_{CD}
-13 \GIJ_{AB} \GIJ_{CD} + {\textstyle {1 \over 4}} \big( h \HB
    \big)_{AB} \big( h \HB  \big)_{CD} \Big\}\,, &(A.12) \cr}
$$
and
$$
\eqalignno{
h_{C[A} h_{B]D} & = {\textstyle {1 \over {32}}}
\Big\{ \half h_{AB} ( h )^3_{CD} +  \half
  \HA_{AB} \big( h \HA h \big)_{CD}
  + {\textstyle {1 \over {4!}}} \big( h \HB \big)_{AB} \big( h
    h \HB h \big)_{CD} \Big\}     \cr
& = {\textstyle {1 \over {32}}} \Big\{ - h_{AB} h_{CD}
    -  \GIJ_{AB} \GIJ_{CD} -
     {\textstyle {1 \over {12}}} \big( h \HB
    \big)_{AB} \big( h \HB \big)_{CD} \Big\} \,. &(A.13)\cr  }
$$
Again, it is easy to check that
$$
\GIJ_{C[A} \GIJ_{B]D} + 3 h_{C[A} h_{B]D}
 = - \half \left( \GIJ_{AB} \GIJ_{CD} + 3 h_{AB} h_{CD} \right) \,,
                        \eqno (A.14)
$$
so the identity (A.1) now holds with $A=3$.

Finally, for $\underline {N=9}$, we have $d=16$. As for $N=16$,
there are no complex structures; a complete and linearly
independent set of real antisymmetric $16\times 16$ matrices is given
by the $9\choose 2$ matrices $\HA$ and the $9\choose 6$ matrices
$\HC$. The relevant Fierz identity now reads
$$
\G^{IJ}_{C[A} \G^{IJ}_{B]D} =
     {\textstyle  {1 \over {16}} {1 \over {2!}} }
   \HA_{AB} \big( \GIJ \HA  \GIJ \big)_{CD} +
 {\textstyle  {1 \over {16}} {1 \over {6!}}}
   \HC_{AB} \big( \GIJ \HC  \GIJ \big)_{CD} \,. \eqno (A.15)
$$
{}From (A.4), we now get
$\GIJ \HA  \GIJ = - 16 \HA $ and, by another fortunate numerical
coincidence, $\GIJ \HC  \GIJ = 0$. Except for the different range of
indices, the resulting identity is the same as (A.7), so (A.1)
is again obeyed with $A=0$.

There is no need at this point to discuss other values of $N$, since
we know from the classification of Lie algebras
that, apart from $G_2$, there are no other exceptional Lie
algebras besides the ones considered above. We have given a
pedestrian and rather explicit construction of these algebras, not
least because, except for $E_8$, the relevant Fierz identities do
not seem to have been discussed anywhere in the literature. From
the present point of view, there exist no exceptional Lie algebras
beyond $E_8$ because the number of terms that must cancel
after the Fierz rearrangements becomes too large, so that
(A.1) can no longer be satisfied.

\bigskip

\centerline{\bf References}
\medskip

\item{[1]} A.~Salam and E.~Sezgin, {\it Supergravities in
Diverse Dimensions}, North-Holland/World Scientific (1989).
\item{[2]} S.~Deser, R.~Jackiw and G.~'t~Hooft, Ann.~Phys. (NY)
{\bf152} (1984) 220; \hfil\break
S.~Deser and R.~Jackiw, Ann.~Phys. {\bf 153} (1984)405; \hfil\break
J.~Abbot, S.~Giddings and K.~Kuchar, Gen.~Rel.~Grav. {\bf 16}
(1984) 751.
\item{[3]} E.~Witten, Nucl.~Phys. {{\bf 311} (1988/89) 46.
\item{[4]} A.~Ashtekar, Phys.~Rev.~Lett. {\bf 57} (1986)
2244; {\it New perspectives in canonical gravity}, Bibliopolis,
1988.
\item{[5]} I.~Bengtsson, Phys. Lett. {\bf B220} (1989) 51.
\item{[6]} A.~Ashtekar, V.~Husain, C.~Rovelli, J.~Samuel
  and L.~Smolin, Class. Quantum Grav. {\bf 6} (1989) L185.
\item{[7]} H.~Nicolai and H.~J.~Matschull, {\it Aspects of Canonical
   Gravity and Supergravity}, preprint DESY 92-099 (1992).
\item{[8]} B.~Julia, in {\it  ``Superspace and Supergravity" },
eds. S.W.~Hawking and M.~Ro\v{c}ek, Cambridge Univ. Press, 1980;
in   {\it ``Unified Field Theories and Beyond"},
Johns Hopkins Workshop on Current Problems in Particle Physics,
Johns Hopkins University, Baltimore, 1981.
\item{[9]} P.~Breitenlohner and D.~Maison, Ann. Inst.
  Poincar\'e {\bf 46} (1987) 215.
\item{[10]} H.~Nicolai, in {\it ``Recent Aspects of Quantum Fields"},
   eds. H.~Mitter and H.~Gausterer, Lecture Notes in Physics 396,
   Springer Verlag, 1991.
\item{[11]} B.~de~Wit, M.T.~Grisaru, E.~Rabinovici and
H.~Nicolai, Phys.~Lett. {\bf B286} (1992) 78.
\item{[12]} N.~Marcus and J.H.~Schwarz, Nucl.~Phys. {\bf
B228} (1983) 145.
\item{[13]} L. Alvarez-Gaum{\'{e} and D.Z. Freedman,
Commun. Math. Phys. {\bf 80} (1981) 443.
\item{[14]} B.~de~Wit and P.~van~Nieuwenhuizen, Nucl.~Phys.
{\bf B312} (1989) 58.
\item{[15]} E. Bergshoeff, E. Sezgin and H. Nishino, Phys. Lett.
    {\bf 186B} (1987) 167.
\item{[16]} H.~Nicolai, Phys. Lett. {\bf 194B} (1987) 402.
\item{[17]} M.A.~Vasiliev, Phys. Lett. {\bf B257} (1991)
111.
\item{[18]} A.L.~Besse, {\it Einstein Manifolds}, Springer, 1987,
Chapter~10.
\item{[19]}
 M.~Berger, Bull.~Soc.~Math.~France {\bf 83} (1955) 279; \hfil \break
  J.~Simons, Ann.~of~Math. {\bf 76} (1962) 213.
\item{[20]} M.F.~Atiyah, R.~Bott and A.~Shapero, Topology
{\bf 3}, Sup.~1 (1964) 3.
\item{[21]} H.~Georgi, {\it Lie algebras in Particle Physics},
Benjamin/Cummings, 1982.
\item{[22]} S.~Okubo, J.~Math.~Phys. {\bf 32} (1991) 1657.
\item{[23]} B.~de~Wit and A.~Van~Proeyen, Commun.~Math.~Phys., in
press.
\item{[24]} C. Chevalley, {\it Theory of Lie Groups}, Princeton
Univ. Press, 1946, page 185.
\item{[25]} K. Yano, {\it Differential geometry on complex and
almost complex spaces}, Pergamon, 1965;\hfil\break
K. Yano and M. Kon, {\it Structures on manifolds}, World Scientific,
1984;\hfil\break
S. Kobayashi and K. Nomizu, {\it Foundations of differential
geometry}, Vol.~2, Wiley, 1969.
\item{[26]}D.V.~Alekseevski\u\i, Funk.~Anal.~i~Prilo\u{z}en
{\bf 2} (1968) 1.
\item{[27]} B.~de~Wit and A.~Van~Proeyen, Nucl.~Phys. {\bf
B245} (1984) 89.
\item{[28]} J.~Bagger and E.~Witten, Nucl.~Phys. {\bf B222}
(1983) 1.
\item{[29]} S.~Ferrara and S.~Sabharwal, Nucl.~Phys. {\bf B332}
(1990) 317.
\item{[30]} R.~Brown and A.~Gray, in {\it Differential Geometry
    in Honor of K.~Yano}, Kinokunya, Tokyo, 1972.
\item{[31]}D.V.~Alekseevski\u\i, Math.~USSR~Izvestija {\bf 9}
(1975) 297.
\item{[32]}  E.~Cremmer and B.~Julia, Nucl.~Phys. {\bf B159}
     (1979) 141;  \hfil\break
    B.~de Wit and H.~Nicolai, Nucl.~Phys. {\bf B208} (1982) 322.
\item{[33]} L.~Castellani, R~D'Auria and P.~Fr\'e, {\it Supergravity
and Superstrings}, Vol.~1, World Scientific, 1991.

\end